 \documentclass{emulateapj}

\shorttitle{An extreme proto-cluster of luminous dusty starbursts at $z_{\rm spec} = 4.002$}
\shortauthors{Oteo et al.}

\usepackage{natbib}
\usepackage{amsmath}
\usepackage{xcolor}
\usepackage{epsfig}
\usepackage{graphicx}

\begin{document}


\title{An extreme proto-cluster of luminous dusty starbursts in the early Universe}


\author{
I.~Oteo\altaffilmark{1,2}, 
R.\,J.~Ivison\altaffilmark{2,1}, 
L.~Dunne\altaffilmark{1,3},
A.~Manilla-Robles\altaffilmark{2,4},
S.~Maddox\altaffilmark{1,3},  
A.\,J.\,R.~Lewis\altaffilmark{1}, 
G.~de~Zotti\altaffilmark{5}, 
M.~Bremer\altaffilmark{6},
D.~L.~Clements\altaffilmark{7}, 
A.~Cooray\altaffilmark{8}, 
H.~Dannerbauer\altaffilmark{9,10},
S.~Eales\altaffilmark{3},
J.~Greenslade\altaffilmark{7}, 
A.~Omont\altaffilmark{11,12}, 
I.~Perez--Fourn\'on\altaffilmark{8,9}, 
D.~Riechers\altaffilmark{13}, 
D.~Scott\altaffilmark{14}, 
P.~van~der~Werf\altaffilmark{15},
A. Weiss\altaffilmark{16}, and
Z-Y.~Zhang\altaffilmark{1,2}
}

\affil{$^1$Institute for Astronomy, University of Edinburgh, Royal Observatory, Blackford Hill, Edinburgh EH9 3HJ}
\affil{$^2$European Southern Observatory, Karl-Schwarzschild-Str. 2, 85748 Garching, Germany}
\affil{$^3$School of Physics and Astronomy, Cardiff University, The Parade, Cardiff CF24 3AA}
\affil{$^4$Department of Physics and Astronomy, University of Canterbury, Private Bag 4800, Christchurch, 8140, New Zealand}
\affil{$^5$INAF, Osservatorio Astronomico di Padova, Vicolo Osservatorio 5, I-35122 Padova, Italy}
\affil{$^{6}$H. H. Wills Physics Laboratory, University of Bristol, Tyndall Avenue, Bristol BS8 1TL}
\affil{$^{7}$Physics Department, Blackett Lab, Imperial College, Prince Consort Road, London SW7 2AZ, UK}
\affil{$^{8}$Department of Physics and Astronomy, University of California, Irvine, CA 92697, USA}
\affil{$^{9}$Instituto de Astrof\'isica de Canarias (IAC), E-38205 La Laguna, Tenerife, Spain}
\affil{$^{10}$Universidad de La Laguna, Dpto. Astrof\'isica, E-38206 La Laguna, Tenerife, Spain}
\affil{$^{11}$CNRS, UMR 7095, Institut d'Astrophysique de Paris, F-75014, Paris, France}
\affil{$^{12}$UPMC Univ. Paris 06, UMR 7095, Institut d'Astrophysique de Paris, F-75014, Paris, France}
\affil{$^{13}$Cornell University, Space Sciences Building, Ithaca, NY 14853, USA}
\affil{$^{14}$Department of Physics and Astronomy, University of British Columbia, Vancouver, BC V6T 1Z1, Canada}
\affil{$^{15}$Leiden Observatory, Leiden University, P.O. Box 9513, NL-2300 RA Leiden, The Netherlands}
\affil{$^{16}$Max-Planck-Institut f\"ur Radioastronomie, Auf dem H\"ugel 69 D-53121 Bonn, Germany}
\email{ivanoteogomez@gmail.com}

\begin{abstract}

We report the identification of an extreme proto-cluster of galaxies in the early Universe whose core (nicknamed Dusty Red Core, DRC, because of its very red color in the {\it Herschel} SPIRE 250-, 350- and 500-$\mu$m bands) is formed by at least ten dusty star-forming galaxies (DSFGs), spectroscopically confirmed to lie at $z_{\rm spec} = 4.002$ via detection of [C\,{\sc i}](1--0), $^{12}$CO(6--5), $^{12}$CO(4--3), $^{12}$CO(2--1) and ${\rm H_2O} (2_{11} - 2_{02})$ emission lines, detected using ALMA and ATCA.  The spectroscopically-confirmed components of the proto-cluster are distributed over a ${\rm 260\, kpc \times 310\, kpc}$ region and have a collective obscured star-formation rate (SFR) of $\sim 6500 \, M_\odot \, {\rm yr}^{-1}$, considerably higher than has been seen before in any proto-cluster of galaxies or over--densities of DSFGs at $z \gtrsim 4$.  Most of the star formation is taking place in luminous DSFGs since no Ly$\alpha$ emitters are detected in the proto-cluster core, apart from a Ly$\alpha$ blob located next to one of the DRC dusty components, extending over $60\,{\rm kpc}$.  The total obscured SFR of the proto-cluster could rise to ${\rm SFR} \sim 14,400 \, M_\odot \, {\rm yr}^{-1}$ if all the members of an over-density of bright DSFGs discovered around DRC in a wide-field LABOCA 870-$\mu$m image are part of the same structure. [C\,{\sc i}](1--0) emission reveals that DRC has a total molecular gas mass of at least $M_{\rm H_2} \sim 6.6 \times 10^{11}\,M_\odot$, and its total halo mass could be as high as $\sim 4.4 \times 10^{13}\,M_\odot$, indicating that it is the likely progenitor of a cluster at least as massive as Coma at $z = 0$. The relatively short gas-depletion times of the DRC components suggest either the presence of a mechanism able to trigger extreme star formation simultaneously in galaxies spread over a few hundred kpc or the presence of gas flows from the cosmic web able to sustain star formation over several hundred million years.


\end{abstract}


\section{Introduction}

Proto-clusters of galaxies are key to tracing the formation of the most massive dark-matter halos in the Universe and represent excellent laboratories in which to confront cosmological simulations \citep{Borgani2011ASL.....4..204B} as well as tools to test and constrain cosmology \citep{Allen2011ARA&A..49..409A,Harrison2012MNRAS.421L..19H}.  Furthermore, whereas the contribution of cluster of galaxies to the cosmic star-formation rate density \citep{Madau2014ARA&A..52..415M} in the local Universe is very low, the contribution of proto-clusters might represent up to $\sim 25$\% at $z \sim 4$ and $\sim 50$\% at $z \sim 10$. In parallel, the fractional cosmic volume occupied by proto-clusters increases by three orders of magnitude from $z \sim 0$ to $z \sim 7$ \citep{Chiang2017ApJ...844L..23C}.  All this highlights the importance of proto-clusters in the early Universe for our understanding of galaxy and structure formation and hierarchical growth \citep{Overzier2016A&ARv..24...14O}.

A ubiquitous feature of galaxy clusters up to $z \sim 2$ is the presence of a strong red sequence produced by massive, passively-evolving early-type galaxies, which dominate their cores \citep[see e.g.][]{Bremer2006MNRAS.371.1427B,Stanford2006ApJ...646L..13S,Rosati2009A&A...508..583R,Hilton2009ApJ...697..436H}. The bulk of their star formation occurred at $z > 2$, over relatively short periods, in a phase compatible with luminous dusty star-forming galaxies \citep[DSFGs -- e.g.][]{Casey2014PhR...541...45C} at $z > 3$--4 \citep{Collins2009Natur.458..603C,Thomas2010MNRAS.404.1775T,Gobat2011A&A...526A.133G,Zeimann2012ApJ...756..115Z}. Consequently, the average total star-formation rate (SFR) in galaxies that go on to populate rich cluster cores might reach several $\times 1000 \, {\rm M}_\odot \, {\rm yr}^{-1}$ \citep[for a Salpeter stellar initial mass function, IMF, though see][]{Romano2017MNRAS.470..401R} for the most massive examples during the early stages. This implies that developing proto-cluster cores should be identifiable as sub-Mpc regions with multiple submm-bright DSFGs with total SFRs of thousands of $M_\odot \, {\rm yr}^{-1}$. Such structures have been already found up to $\sim 3$ \citep[e.g.][]{Ivison2013ApJ...772..137I,Dannerbauer2014A&A...570A..55D,Yuan2014ApJ...795L..20Y,Umehata2014MNRAS.440.3462U,Umehata2015ApJ...815L...8U,Casey2015ApJ...808L..33C,Flores-Cacho2016A&A...585A..54F,Planck2016A&A...596A.100P,Wang2016ApJ...828...56W,Hung2016ApJ...826..130H} but those reported at $z \gtrsim 4$ are much less extreme and typically contain only one DSFG \citep{Daddi2009ApJ...694.1517D,Capak2011Natur.470..233C,Walter2012Natur.486..233W}.

As a result of our systematic search for ultrared DSFGs \citep[sources whose SPIRE flux densities increase from $250 \, {\rm \mu m}$ to $500 \, {\rm \mu m}$ --][]{Ivison2016ApJ...832...78I} in the {\it H}-ATLAS survey \citep{Eales2010PASP..122..499E}, we discovered one system with exactly those characteristics, which is therefore the ideal observational model of the early evolution of proto-clusters.  This source was nicknamed the Dusty Red Core (DRC, R.A. = 00:42:23.8, decl. = $-$33:43:34.8), and it might represent the core of a larger proto-cluster of galaxies at $z_{\rm spec} = 4.002$.  This nickname reflects the fact that this was the reddest source found in {\it H}-ATLAS with $S_{250} < S_{350} < S_{500} < S_{870}$. Consequently, it was initially thought to lie at very high redshift, $z \sim 5$--6, according to its photometric redshift \citep{Ivison2016ApJ...832...78I}.  For this reason, most of the initial follow-up was designed to detect CO lines at those redshifts.

In this paper we report on the nature of this unique system, along with the observations which led to its discovery.  The paper is organized as follows: \S\ref{section_observations} describes the observations used in this work.  \S\ref{section_discovery_DRC} reports the discovery of this unique system, and then in \S\ref{section_properties_DRC} we discuss its physical properties, including SFR, molecular gas mass, total mass (baryonic and dark matter) and gas-depletion time.  Finally, \S\ref{conclusions_of_the_paper} presents the main conclusions of the paper.  The total IR luminosities ($L_{\rm IR}$) reported in this work refer to the integrated luminosities between rest-frame 8 and $1000\,{\rm \mu m}$, and the SFRs are derived from $L_{\rm IR}$ by assuming a Salpeter IMF and the classical \cite{Kennicutt1998ARA&A..36..189K} calibration.  Throughout the paper, we assume a $\Lambda$CDM cosmology with $H_0=70 \, {\rm km \,s^{-1}\, Mpc^{-1}}$, $\Omega_\Lambda = 0.7$ and $\Omega_{\rm M} = 0.3$.


\section{Observations}
\label{section_observations}

\subsection{APEX observations}

The Atacama Pathfinder Experiment (APEX) telescope's Large APEX BOlometer CAmera (LABOCA -- \citealt{Kreysa2003SPIE.4855...41K,Siringo2009A&A...497..945S}) $870 \, {\rm \mu m}$ observations were carried out during 2013 October under projects M-092.F-0015-2013 (P.I.\ A. Weiss) and 191A-0748 (P.I.\ R.\,J.~Ivison). A compact-raster scanning mode was used, whereby the telescope scans in an Archimedean spiral for $t_{\textrm{int}}=35\,\textrm{sec}$ at four equally spaced raster positions in a $27''\times27''$ grid. Each scan was approximately $t_{\textrm{int}}\approx7\,\textrm{min}$ long such that each raster position was visited three times leading to a fully sampled map over the full $11\,'$-diameter field of view of LABOCA.  During the observations, we recorded typical precipitable water vapour (PWV) values between 0.4--1.3\,mm, corresponding to a zenith atmospheric opacity of $\tau=0.2$--0.4. Finally, the flux density scale was determined to an r.m.s.\ accuracy of $\sigma_{\textrm{calib}}\approx7$\% using observations of the primary calibrators, Uranus and Neptune, whilst pointing was checked every hour using nearby quasars and found to be stable to $\sigma_{ \textrm{point}}\approx3''$ (r.m.s.).  A total of $11.4 \, {\rm h}$ were spent integrating on our target, covering an area of $124 \, {\rm arcmin}^2$. The final map was then beam-smoothed, to a resolution of $27''$. The average r.m.s.\ background noise is then $\sim 1.9 \, {\rm mJy \, beam^{-1}}$. The data were reduced using the Python-based BOlometer data Analysis Software package \citep[BOA \textsc{v4.1} ---][]{Schuller2012SPIE.8452E..1TS}, following the prescription outlined in \cite{Siringo2009A&A...497..945S} and \cite{Schuller2009A&A...504..415S}.  More details about the observations, data reduction and source extraction can be found in Lewis et al.\ (in prep).

\subsection{ALMA observations}\label{section_ALMA_data}

The ALMA data presented in this paper come from four different projects: 2013.1.00449.S (P.I.\ A.~Conley), 2013.A.00014.S -- a Director's Discretionary Time (DDT) proposal (P.I.\ R.\,J.~Ivison), 2013.1.00001.S (P.I.\ R.\,J.~Ivison) and 2016.1.01287.S (P.I.\ I.~Oteo). 

In project 2013.1.00449.S we carried out spectral scans in the 3-mm band on a sample of eight ultra-red starbursts at $z_{\rm phot} \gtrsim 4$ selected from HerMES \citep{Oliver2012MNRAS.424.1614O} and {\it H}-ATLAS surveys with the aim of measuring their redshift via multiple CO line detections \citep{Oteo2016ApJ...827...34O,Asboth2016MNRAS.462.1989A,Fudamoto2017arXiv170708967F, Riechers2017arXiv170509660R}. The observations were taken between 2014 July 03 and August 28, in a total of five scheduling blocks (SBs) corresponding to the five tunings needed to cover most of ALMA band 3, between 84 and $114.88\,{\rm GHz}$. The data for each SB were reduced in {\sc CASA}, following standard procedures. Imaging was carried out using natural weighting to improve sensitivity, resulting in an average r.m.s.\ sensitivity of $\sim 0.75 \, {\rm mJy \, beam^{-1}}$ in channels binned to $100 \, {\rm km \, s^{-1}}$.  The {\sc fwhm} synthesized beam ranged between $0.6''$ and $1.2''$ due to the different array configurations used in the different tunings.

In project 2013.A.00014.S we observed DRC for about one hour, aiming to confirm what were thought to be two faint emission lines, $^{12}$CO(7--6) and [C\,{\sc i}] (2--1), detected in DRC at around $98.4\,{\rm GHz}$ in data from the earlier project, 2013.1.00449.S\footnote{As we see later, these two putative lines were revealed to be one extremely broad line, namely [C\,{\sc i}] (1-0).}. The observations were carried out on 2015 January 14, when the array was in a compact configuration.  The data were reduced using the ALMA pipeline and imaged using natural weighting to improve sensitivity.  The resulting beam size was $\sim 2.0''$ {\sc fwhm} and the r.m.s.\ sensitivity was $\sim 0.13 \, {\rm mJy \, beam^{-1}}$ in $100 \, {\rm km \, s^{-1}}$ channels.  

Project 2013.A.00014.S consisted of high-spatial-resolution ($0.12''$) continuum observations of the brightest DRC component, what we will call DRC-1, at $870 \, {\rm \mu m}$.  The aim of the observations was to study the morphology and extent of the dust emission in a sub-sample of the ultrared DSFGs presented in \cite{Ivison2016ApJ...832...78I}.  Details on the observations, data calibration and imaging can be found in \cite{Oteo2016ApJ...827...34O} and \cite{Oteo2017arXiv170904191O}. Briefly, the data were calibrated using the ALMA pipeline and imaging was done using Briggs weighting, which represents a good compromise between depth and spatial resolution. The resulting r.m.s.\ sensitivity is $\sim 0.1 \, {\rm mJy \, beam^{-1}}$ with a {\sc fwhm} synthesized beam of $\sim 0.12''$ or $\sim 830 \, {\rm pc}$ at $z = 4.002$.

In project 2016.1.01287.S we observed DRC at $2 \, {\rm mm}$ in 14 SBs, each about $70 \, {\rm min}$ long, with the aim of detecting one or more additional emission lines to determine the redshift of DRC unambiguously.  Due to the smaller primary beam in band 4, a two-pointing mosaic was used, such that roughly the same area was covered as for the earlier 3-mm observations, without significant primary beam attenuation.  The data were calibrated using the ALMA pipeline and imaging was done using natural weighting in the concatenated visibilities corresponding to the two pointings.  The resulting r.m.s.\ sensitivity is $\sim 50 \, {\rm \mu Jy \, beam^{-1}}$ in $100 \, {\rm km \, s^{-1}}$ channels and $\sim 6\,{\rm \mu Jy\, beam^{-1}}$ in the continuum map, with a synthesized beam size ({\sc fwhm}) of $1.68'' \times 1.54''$ in both pointings.

\subsection{Jansky VLA observations}

We used the Karl G.\ Jansky Very Large Array (VLA) to observe DRC, covering 27.68--28.58\,GHz and 32.01--32.90\,GHz using the 8-bit samples, in the CnB configuration, during 2015 January 24--25 (project VLA/14B-497; P.I.\ R.\,J.~Ivison). The data were calibrated using the VLA pipeline, and the calibrated visibilities were imaged using natural weighting to improve the sensitivity.  These observations were aimed at detecting low-$J$ CO lines from DRC, but after the redshift was confirmed with the ALMA and ATCA observations, we knew that no CO lines were covered by the VLA spectral setup.  Therefore, we utilize only the continuum map in this paper, which has an r.m.s.\ sensitivity of $\sim 6.3 \, {\rm \mu Jy \, beam^{-1}}$ and a synthesized beam ({\sc fwhm}) of $1.07'' \times 0.66''$.

\begin{figure*}
\centering
\includegraphics[width=\textwidth]{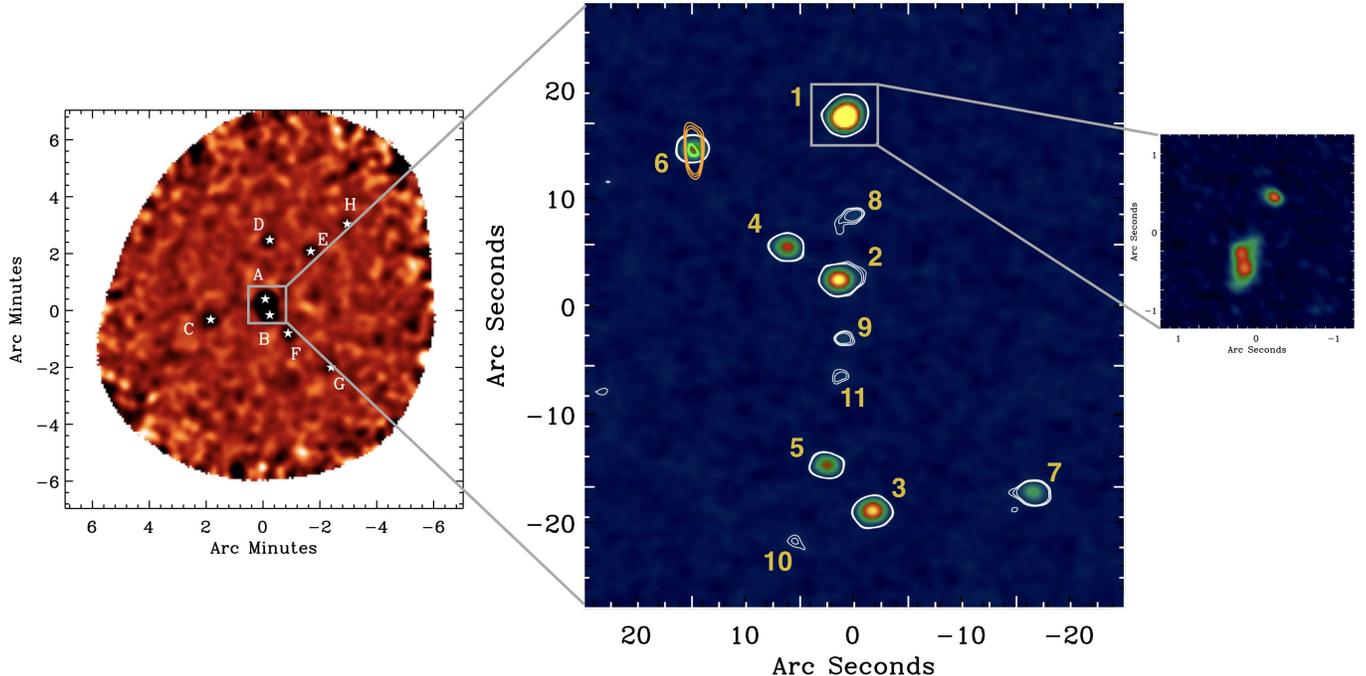} 
\caption{DRC from large to small scales -- the core of an extreme proto-cluster of galaxies at $z_{\rm spec} = 4.002$. The left panel shows a wide-field LABOCA map at $870 \, {\rm \mu m}$ of the environment of DRC, smoothed to a resolution of $27''$.  The eight DSFGs suggest an over-dense field, because we would expect $2 \times$ times less sources (Lewis et al., in prep) according to the most recent single-dish submm number counts \citep[e.g.][]{Geach2017MNRAS.465.1789G}.  DRC is the extended LABOCA source associated with the DSFGs labelled  A and B. The DSFGs around DRC (labelled C--H) all have $5 < S_{\rm 870 \, \mu m}\,[{\rm mJy}] < 11$ and if all of them lie at the same redshift as DRC then the collective obscured SFR would be $\sim 14,400 \, {\rm M}_\odot \, {\rm yr}^{-1}$.  This is considerably higher than any other starbursting structure at $z \gtrsim 4$ reported to date.  The middle panel shows the ultra-deep ALMA $2 \, {\rm mm}$ continuum map of DRC.  Green and orange contours (from $5 \sigma$) represent the radio continuum emission observed by the Jansky VLA and ATCA, respectively.  DRC is resolved into at least 11 components, which is also a significant over-density of DSFGs according to the most recent ALMA number counts \citep{Oteo2016ApJ...822...36O,Aravena2016ApJ...833...68A,Dunlop2017MNRAS.466..861D}. The right panel shows a high-resolution ALMA continuum map at $870 \, {\rm \mu m}$ of the brightest DSFG in the proto-cluster, referred to here as DRC-1.  This component is resolved into at least three star-forming clumps, whose interaction likely produces the extremely broad [C\,{\sc i}](1--0) and $^{12}$CO(6--5) line profiles (see Fig.~\ref{DRC_spec_components_CO65}).}
\label{figure_LABOCA_wide_map}
\end{figure*}

\subsection{VLT observations}

DRC was observed on the nights of 2015 August 13, September 5 and 7 with the Multi Unit Spectroscopic Explorer (MUSE) integral field spectrograph mounted on European Southern Observatory's Very Large Telescope UT4 \citep{Bacon2010SPIE.7735E..08B} (DDT program 295.A-5029, P.I.\ I.~Oteo), with a seeing varying between $0.9--1.1''$.  The $1' \times 1'$ MUSE field of view was centered on the proto-cluster core, covering the 11 components detected in our ALMA Cycle 2 and DDT programs.  With the nominal wavelength range (475--930\,nm), we performed a series of exposures, each $15 \, {\rm min}$, for a total on-source time of $3 \, {\rm h}$.  Between individual exposures, the spectrograph was rotated by 90 degrees and a random dithering pattern was added.  The data were reduced using the MUSE pipeline, which performed all the basic reduction steps \citep{Bacon2015A&A...575A..75B}.  The subtraction of the sky emission was improved in the reduced cubes using a set of custom scripts.

FORS2 broad-band imaging in the $I$ filter were carried out during $4\, {\rm h}$ on the night of 2014 December 17 (project 093.A-0705, P.I: R.~J.~Ivison) under good weather conditions with seeing $\sim 1''$.  The data were reduced with the {\sc esorex} pipeline following the standard procedures and the final images were astrometrically calibrated using the available VIKING $z$-band imaging in the field.  The observations reached a $5 \sigma$ limiting magnitude of $25.3 \,{\rm mag}$.

\subsection{Gemini observations}

Broad-band near-IR observations in the $K_s$ filter were carried out between 2014 18 of July 18 and November 04 with FLAMINGOS-2 mounted in the Gemini-South telescope (project GS-2014A-Q-58, P.I.\ L.~Dunne).  DRC was observed for a total of $4.1 \, {\rm h}$ with an average seeing of $0.72''$. A classical dither pattern was used to remove the sky emission during the data reduction, performed with THELI \citep{Schirmer2013ApJS..209...21S}. First, the flat-field and dark correction was applied to each science frame, then the sky emission in each science frame was subtracted using a dynamical model of four images taken immediately before and after each frame.  The flat-field correction was applied, cosmic rays were removed and the vignetted region of the image was masked.  Finally, the images were combined using SWarp \citep{Bertin2010ascl.soft10068B} and astrometrically corrected and flux calibrated using 2MASS \citep{Skrutskie2006AJ....131.1163S}.  The final image has a $3 \, \sigma$ limiting magnitude of $25.03 \, {\rm mag}$ in a $2''$ aperture.

\subsection{Spitzer/IRAC}

DRC was observed with {\em Spitzer}/IRAC at 3.6 and 4.5\,$\mu$m on 14 September 2015 (program ID:\ 11107; PI:\ P\'erez-Fournon).
A 36-position dither pattern with 30 second exposures per frame was used, 
totaling 1080 second integrations in each band. Data reduction was performed with the {\em MOPEX} 
package using standard procedures. Absolute astrometry was obtained relative to Gaia DR1, yielding rms 
accuracies of 0.04$''$ and 0.06$''$ in the 3.6 and 4.5\,$\mu$m bands, respectively.

\subsection{ATCA observations}

The $^{12}$CO(2--1) emission from DRC was observed with the Australia Telescope Compact Array (project C3185, P.I.\ I.~Oteo) during 2017 July 18--19 in its most compact configuration, with the $64 \, {\rm MHz}$ spectral mode.  The set-up of the observations was carried out using the calibrator 1921$-$293, which was also used for bandpass calibration.  The absolute flux scale (which we estimate to have an uncertainty of $\sim 15\,\%$) was determined by observing 1934$-$638, while 0104$-$408 was used as the phase calibrator.  The observations were carried out in good weather conditions and covered almost full tracks, for a total on-source time of $\sim 14 \, {\rm h}$.  The data were calibrated by using the standard techniques in MIRIAD, including manual flagging of bad data (note that antenna CA06 was used in the observations but these data were flagged after the calibration was complete due to the poor data quality with respect to the other antennas).  The calibrated visibilities were then transformed into CASA format, where the cubes and continuum maps were created by using natural weighting to improve sensitivity. The resulting synthesized beam {\sc fwhm} is $14.2'' \times 10.6''$ and the r.m.s.\ sensitivity is $\sim 0.13 \, {\rm mJy \, beam^{-1}}$ in $\sim 800 \, {\rm km \, s^{-1}}$ wide channels.

Continuum observations with ATCA at $32.5$, $9.0$, and $5.5 \, {\rm GHz}$ were carried out between 2013 October and 2015 July (project C2905, P.I.\ L.~Dunne).  The $7 \, {\rm mm}$ observations were initially aimed at confirming the redshift of DRC via detection of $^{12}$CO, but no lines were detected and thus we only use the continuum maps here.  The data were reduced in MIRIAD following standard procedures, then the continuum maps were created from the calibrated visibilities in {\sc CASA}. At $32.5 \, {\rm GHz}$ the continuum map reached an r.m.s.\ sensitivity of $\sim 7.6 \, {\rm \mu Jy \, {\rm beam}^{-1}}$ with a synthesized beam ({\sc fwhm}) of $1.3'' \times 0.7''$. The observations at $9.0$ and $5.5 \, {\rm GHz}$ reached r.m.s.\ sensitivities of $\sim 6.6 \, {\rm \mu Jy \, {\rm beam}^{-1}}$.

\section{An extreme over-density of DSFGs}\label{section_discovery_DRC}

The wide-field LABOCA map at $870\,{\rm \mu m}$ (see left panel of Figure \ref{figure_LABOCA_wide_map}) revealed a $2 \sigma$ over-density (Lewis et al., in prep) of DSFGs with respect to the most recent single-dish number counts at $870\,{\rm \mu m}$ \citep{Geach2017MNRAS.465.1789G}. The brightest DSFG of this over-density (components A and B in the left panel of Fig.~\ref{figure_LABOCA_wide_map}, which form an extended $870 \, {\rm \mu m}$ source) was followed-up with ALMA in the $2 \, {\rm mm}$ and $3 \, {\rm mm}$ bands, aiming to determine a precise, unambiguous redshift via detection of several CO lines, similarly to the successful redshift determinations obtained by \citet{Weiss2009ApJ...705L..45W,Cox2011ApJ...740...63C,Weiss2013ApJ...767...88W,Riechers2013Natur.496..329R,Asboth2016MNRAS.462.1989A,Strandet2016ApJ...822...80S,Oteo2016ApJ...827...34O,Fudamoto2017arXiv170708967F,Riechers2017arXiv170509660R}.

When our first deep 3-mm ALMA observations (see \S \ref{section_ALMA_data}) were delivered, we found that the flux density at $870 \, {\rm \mu m}$ (seen as extended emission in the LABOCA map of the source, see left panel of Figure \ref{figure_LABOCA_wide_map}) was caused by an accumulation of bright DSFGs at $z =3-5$ instead of being associated to a single source or merging system at very high redshift.

The ultra-deep ALMA continuum map at $2\,{\rm mm}$ reveals that the extended LABOCA source associated with DRC is resolved into at least 11 DSFGs (see middle panel of Fig.~\ref{figure_LABOCA_wide_map} and Table \ref{table_properties_DRC_components}).  Six of these components are also detected in the deep continuum map at 3\,mm, while the other five are not detected due to the poorer sensitivity to dust at 3\,mm.  The number of $2 \, {\rm mm}$ sources represents a significant over-density of DSFGs when compared with the most recent ALMA counts \citep{Oteo2016ApJ...822...36O,Aravena2016ApJ...833...68A,Dunlop2017MNRAS.466..861D} because only 1-2 DSFGs brighter than the faintest DRC component would be expected, and we have 11 of them.  For this calculation we have converted the $2 \, {\rm mm}$ flux densities of DRC components into flux densities at $1.2 \, {\rm mm}$ by using the ALESS template \citep{Swinbank2014MNRAS.438.1267S}, which represents the average FIR SED of the classical DSFG population at $z \sim 2.5$ and has been found to provide a good representation of the FIR SED of ultrared DSFGs \citep{Ivison2016ApJ...832...78I}.  This number of components represent a lower limit if we consider that some of the individual $2 \, {\rm mm}$ sources might be resolved into several components if they were observed at higher spatial resolution.  This is what happens for DRC-1, the brightest component of the proto-cluster.  Our high-resolution ALMA imaging at $870 \, {\rm \mu m}$ reveals that this source is resolved into at least three bright star-forming clumps with extreme SFR densities of $\Sigma_{\rm SFR} \sim 800$--$2,000 \, {\rm M}_\odot \, {\rm yr}^{-1} \, {\rm kpc}^{-2}$ (Oteo et al., in prep.), which is among the highest known \citep{Riechers2014ApJ...796...84R,Oteo2017ApJ...837..182O,Iono2016ApJ...829L..10I,Riechers2017arXiv170509660R}.

\begin{table*}[!t]
\caption{\label{table_properties_DRC_components}Properties of DRC components}
\centering
\begin{tabular}{lccccccccc}
\hline
 Component & R.A. & Dec. & $S_{\rm 2\,mm}$& $S_{\rm 3\,mm}$ & $v_{\rm center}$\tablenotemark{(a)} & ${L_{\rm IR}}$ & ${\rm SFR}$\tablenotemark{(b)} 	&	$M_{\rm H_2}$\tablenotemark{(c)}		&	$\tau_{\rm dep}$	\\
  &  &  & $[{\rm mJy}]$ & $[{\rm mJy}]$ & ${\rm km \, s^{-1}}$ & $[\times 10^{11}\,L_\odot]$ & $[{\rm M}_\odot \, {\rm yr}^{-1}]$ 	&	[$\times 10^{11} \, {\rm M}_\odot$]	&	$[{\rm Myr}]$	\\
\hline
DRC--1		&	00:42:23.52	&	$-$33:43:23.4	&	$2.117 \pm 0.058$	&	$0.406 \pm 0.028$		&	$-58 \pm 32$		&	$161.5$		&	$\sim 2900$	&	$\sim 2.62$	&	$\sim 90$		\\	
DRC--2		&	00:42:23.56	&	$-$33:43:38.5	&	$0.723 \pm 0.011$	&	$0.154 \pm 0.010$		&	$470 \pm 97$		&	$55.2$		&	$\sim 990$	&	$\sim 1.18$	&	$\sim 120$		\\	
DRC--3		&	00:42:23.31	&	$-$33:43:59.9	&	$0.659 \pm 0.010$	&	$0.218 \pm 0.022$		&	$286 \pm 12$		&	$50.9$		& 	$\sim 902$	&	$\sim 1.78$	&	$\sim 200$		\\	
DRC--4		&	00:42:23.95	&	$-$33:43:35.4	&	$0.347 \pm 0.099$	&	$0.075 \pm 0.017$		&	$495 \pm 27$		&	$26.5$		&	$\sim 475$	&	$\sim 1.08$	&	$\sim 230$		\\	
DRC--5		&	00:42:23.65	&	$-$33:43:55.7	&	$0.295 \pm 0.094$	&	$0.110 \pm 0.012$		&	--				& 	$22.5$		&	$\sim 404$	&	--			&	--			\\	
DRC--6		&	00:42:24.64	&	$-$33:43:26.4	&	$0.282 \pm 0.065$	&	$0.102 \pm 0.011$		&	$-77 \pm 26$		& 	$21.5$		&	$\sim 386$	&	-			&	-			\\	
DRC--7		&	00:42:22.12	&	$-$33:43:58.2	&	$0.176 \pm 0.082$	&	--					&	$2010 \pm 261$	&	$13.4$		&	$\sim 241$	&	--			&	--			\\	
DRC--8		&	00:42:23.46	&	$-$33:43:32.5	&	$0.055 \pm 0.010$	&	--					&	$-401 \pm 38$		&	$4.2$		&	$\sim 75$		&	--			&	--			\\ 	
DRC--9		&	00:42:23.56	&	$-$33.43.47.3	&	$0.042 \pm 0.011$	&	--					&	$289 \pm 17$		&	$3.2$		&	$\sim 57$		&	--			&	--			\\ 	
DRC--10		&	00:42:23.53	&	$-$33:43:43.9	&	$0.040 \pm 0.007$	&	--					&	$1643 \pm 32$		&	$3.1$		&	$\sim 55$		&	--			&	--			\\
DRC--11		&	00:42:23.87	&	$-$33.44.02.9	&	$0.039 \pm 0.009$	&	--					&	$492 \pm 35$		&	$3.0$		&	$\sim 53$		&	--			&	--			\\
\hline
\tablenotetext{1}{Velocity center of the $^{12}$CO(6--5) emission derived from a Gaussian fit to the line profile.}
\tablenotetext{2}{SFRs have been obtained by re-scaling the ALESS template to the observed $2 \, {\rm mm}$ photometry of the DRC components.  If the Arp\,220 template had been used, the derived SFR would be higher than those presented in this table by a factor of $\times 1.4$.}
\tablenotetext{3}{In this work we only report the molecular gas mass for those components detected in [C\,{\sc i }](1--0); the molecular gas mass derived from $^{12}$CO(6--5) would be highly uncertain due to the need to assume the shape of the CO spectral-line energy distribution (SLED) and the $\alpha_{\rm CO}$ conversion factor \citep[see][]{Ivison2011MNRAS.412.1913I}.}
\end{tabular}
\end{table*}

\begin{figure*}
\centering
\includegraphics[width=0.85\textwidth]{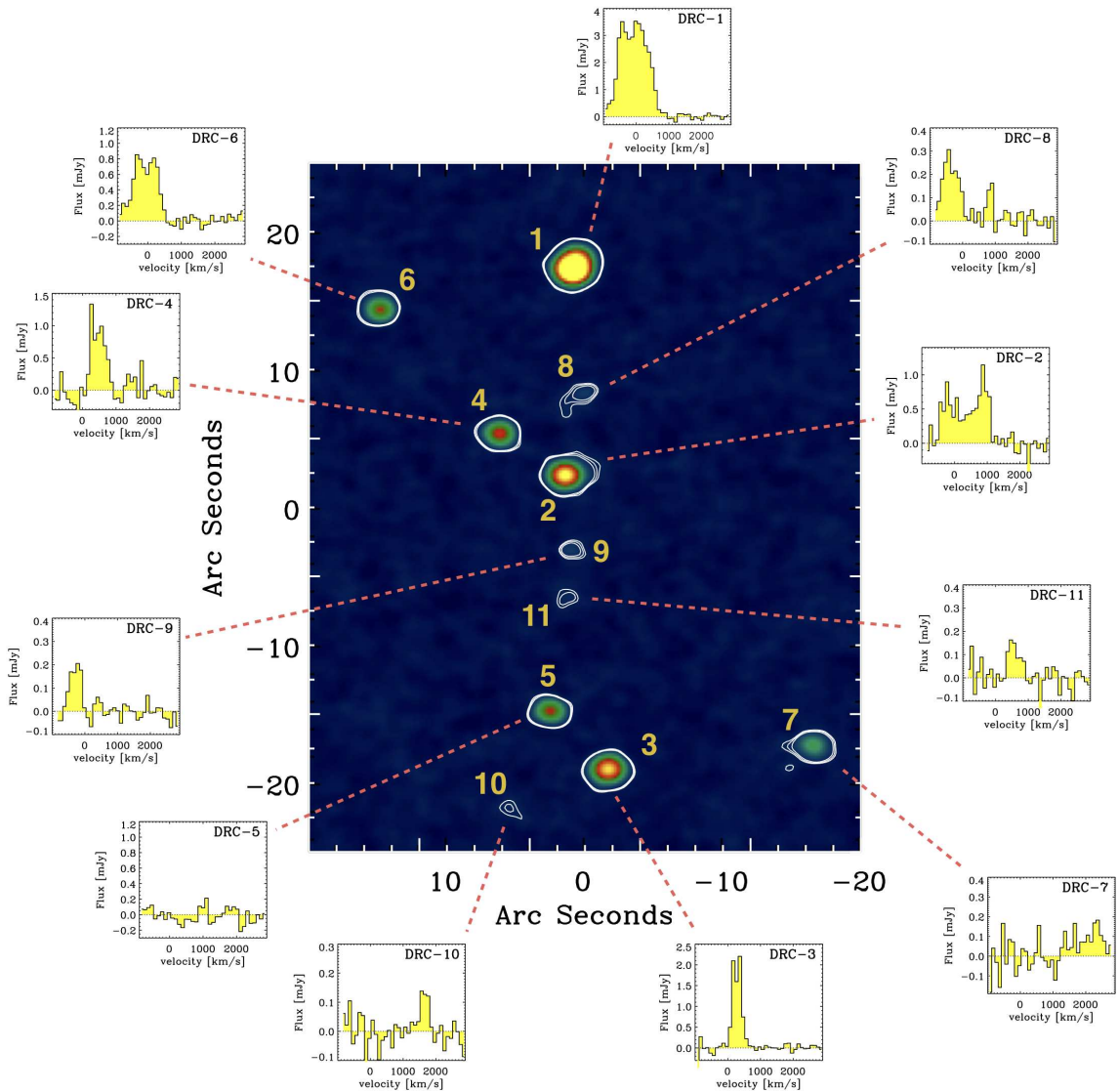} 
\caption{Spectroscopic confirmation of the proto-cluster members at $z_{\rm spec} = 4.002$ via detection of $^{12}$CO(6--5) emission (additional [C\,{\sc i}](1--0), ${\rm H_2O} (2_{11} - 2_{02})$, $^{12}$CO(4--3) and $^{12}$CO(2--1) detections are shown in Figs~\ref{ALMA_spec_DRC_DDT}, \ref{DRC_map_color_IRAC} and \ref{DRC_Lya_image}).  We have considered that the redshift of the structure corresponds to the redshift of DRC-1, its brightest component.  The central panel shows the same $2 \, {\rm mm}$ ultra-deep continuum imaging already shown in the central panel of Fig.~\ref{figure_LABOCA_wide_map}.  The $^{12}$CO(6--5) detections confirm that at least 10 our of the 11 continuum sources belong to the proto-cluster.  The lack of $^{12}$CO(6--5) emission in DRC-5 does not necessarily mean that it does not belong to the proto-cluster, because the spectral setup of our ALMA observations does not cover velocities, $v < -1000 \, {\rm km \, s^{-1}}$.  There is a wide range of line profiles, from relatively narrow emission (DRC-10) to extremely broad lines, as wide as $\sim 1000 \, {\rm km \, s^{-1}}$ (DRC-1) or $\sim 2100 \, {\rm km \, s^{-1}}$ (DRC-2). Note that the horizontal axes are not centered at $v = 0 \, {\rm km \, s^{-1}}$ because the spectral set-up of the ALMA observations only covers $v > -1000 \, {\rm km \, s^{-1}}$.}
\label{DRC_spec_components_CO65}
\end{figure*}

The spectral cubes associated with our ALMA $2 \, {\rm mm}$ and $3 \, {\rm mm}$ observations confirm that at least 10 of these components are at the same redshift, $z_{\rm spec} = 4.002$, via detection of up to five emission lines (we show the $^{12}$CO(6--5) detections in Fig.~\ref{DRC_spec_components_CO65}, the [C\,{\sc i}](1--0) and ${\rm H_2O} (2_{11} - 2_{02})$ detections in Fig.~\ref{ALMA_spec_DRC_DDT}, and the $^{12}$CO(4--3) and $^{12}$CO(2--1) moment-0 maps in Fig.~\ref{DRC_map_color_IRAC} -- see also Table \ref{table_properties_DRC_components_line_fluxes}). They are therefore physically related, belonging to the same, massive structure.  The spectroscopically-confirmed sources extend over an area of ${\rm 260 \, kpc \times 310 \, kpc}$.  This area is smaller than expected for proto-cluster at $z \sim 4$ according to simulations \citep{Chiang2013ApJ...779..127C,Muldrew2015MNRAS.452.2528M}, but relatively close to the expected size of proto-cluster cores at $z = 4$ \citep{Chiang2017ApJ...844L..23C}.  We thus consider it possible that DRC is the core of a larger proto-cluster structure to which at least some of the LABOCA-detected DSFGs around DRC (labelled from C to H in the left panel of Figure \ref{figure_LABOCA_wide_map}) might belong to.   This needs to be confirmed with future spectroscopic observations.  To establish a comparison, the proto-cluster SSA22 at $z = 3.09$ has eight spectroscopically-confirmed DSFGs over an area of $0.7 \, {\rm Mpc} \times 1.4 \, {\rm Mpc}$ \citep{Umehata2015ApJ...815L...8U} while DRC has six spectroscopically-confirmed DSFGs (considering only the sources within the same SFR range) over only ${\rm 260 \, kpc \times 310 \, kpc}$, and thus represents a much more over-dense environment (with DRC having a $10 \times$ times higher source density) in terms of the dusty galaxy population. The same is true of other proto-cluster candidates, whose dusty components are distributed over much larger areas than those in DRC \citep[e.g.][]{Clements2016MNRAS.461.1719C,Flores-Cacho2016A&A...585A..54F}.

Our deep Jansky VLA and ATCA observations revealed continuum emission from DRC-6 (see Fig.~\ref{figure_LABOCA_wide_map}) from $28$ to $5.5 \, {\rm GHz}$ ($S_{\rm 28\, {\rm GHz}} = 96 \pm 15 \, {\rm \mu Jy}$, $S_{\rm 9.0\, {\rm GHz}} = 120 \pm 8 \, {\rm \mu Jy}$ and $S_{\rm 5.5\, {\rm GHz}} = 128 \pm 6 \, {\rm \mu Jy}$).  For a DSFG, its radio spectrum is relatively flat \citep[e.g.][]{Ibar2010MNRAS.401L..53I,Murphy2017ApJ...839...35M}.  Furthermore, the elevated FIR-to-radio flux density ratio in this component -- the radio emission is $\sim 50 \times$ brighter than would be predicted from the ALESS template \citep{Swinbank2014MNRAS.438.1267S} used in several discussions in this paper -- suggests the presence of an AGN \citep{Guidetti2017MNRAS.471..210G}. No radio emission was detected in any of the other DRC components, which could suggest that the radio-loud phase is shorter than the starburst phase. Alternatively, several DRC components could be in a radio-loud phase, hidden from us because their jets are misaligned with our line of sight.  Either way, our radio observations reveal at least one AGN, and thus presumably a black hole that is growing while its host galaxy forms stars.

The existence of a significant over-density of luminous DSFGs at $z_{\rm spec} = 4.002$ over a ${\rm 260 \, kpc \times 310 \, kpc}$ area supports the idea that, at least in some cases, bright DSFGs trace dense environments, which agrees with several previous findings \citep[e.g.][]{Stevens2003Natur.425..264S,Venemans2007A&A...461..823V,Tamura2009Natur.459...61T,Ivison2013ApJ...772..137I,Dannerbauer2014A&A...570A..55D,Umehata2014MNRAS.440.3462U,Umehata2015ApJ...815L...8U,Casey2015ApJ...808L..33C,Wang2016ApJ...828...56W}. 

The nature of DRC was established by following up with ALMA a sample of ultrared DSFGs which were initially selected because of their red SPIRE colors, as confirmed by ground-based imaging observations at 850 or 870\,$\mu$m using SCUBA-2 and/or LABOCA, data which improved the accuracy of photometric redshift estimates \citep{Ivison2016ApJ...832...78I}. These ultrared DSFGs were distributed over all the {\it H}-ATLAS fields, a total sky area of $\sim 600 \, {\rm deg}^{2}$.  The fact that this is the first extreme proto-cluster in {\it H}-ATLAS suggests the number density of such structures is $N[{\rm deg}^{-2}] \gtrsim 1.7 \times 10^{-3}$.  Recently, \cite{Negrello2017MNRAS.470.2253N} presented the expected number of proto-clusters as a function of their total $L_{\rm IR}$ and redshift, using the galaxy evolution models of \cite{Cai2013ApJ...768...21C}.  At $z > 4$, \cite{Negrello2017MNRAS.470.2253N} predicted $N[{\rm deg}^{-2}] > 10^{-2}$ proto-clusters with $L_{\rm IR} \sim 3.7 \times 10^{13}\,L_\odot$, the total IR luminosity of DRC.  These predictions are consistent with DRC because: (1) we can only derive a lower limit on the number of luminous proto-clusters since we have not followed-up with ALMA all ultrared DSFGs in {\it H}-ATLAS, and (2) we have only imaged the central part of the proto-cluster; if some of the LABOCA sources around DRC belong to the same structure then $L_{\rm IR}$ may increase considerably.

\subsection{Redshift confirmation of DRC: the full story}

As an example of the difficulties sometimes encountered when trying to confirm redshifts via the detection of atomic and/or molecular lines in submm and mm windows, it is worth mentioning that the redshift of DRC was only confirmed unambiguously after a large series of observations spanning several years.  First, DRC was included in the sample of ultrared galaxies from the {\it H}-ATLAS survey \citep{Ivison2016ApJ...832...78I,Fudamoto2017arXiv170708967F} selected for spectral scans in the $3 \, {\rm mm}$ band with ALMA.  In those spectral scans of DRC, we detected only very faint emission in the center of the band, which was interpreted at $^{12}$CO(7--6) alongside C\,{\sc i}(2--1).  DRC was then observed through an ALMA DDT program to confirm these lines. However, the observations revealed a single, extremely broad emission line, which we initially associated with CO, which then yielded several discrete possibilities for the redshift. We checked three different redshift options using the Jansky VLA, but no CO line was detected.  Our ALMA Cycle~4 observations, aimed at confirming the redshift of DRC unambiguously, finally detected a second and a third emission line, neither at the frequency we were expecting.  These detections, of what could only be $^{12}$CO(6--5) and ${\rm H_2O} (2_{11} - 2_{02})$, revealed that the initial line detected in the ALMA $3 \, {\rm mm}$ and DDT observations was not CO, but instead [C\,{\sc i}](1--0), and that the redshift of DRC is $z_{\rm spec} = 4.002$.  At this redshift, the $^{12}$CO(4--3) transition was covered by our $3 \, {\rm mm}$ spectral scan, but in a noisy region that prevented line profile being detected.  It should be also noted that because DRC lies close to the edge of the {\it H}-ATLAS image of the South Galactic Pole, it sometimes dropped out of the {\it H}-ATLAS catalogues, depending on how much padding was adopted, leading to some confusion.

\section{Properties of DRC}
\label{section_properties_DRC}

\subsection{Total star-formation rate}
\label{section_SFR_DRC}

In order to estimate the total SFR of the 10 spectroscopically-confirmed DRC components (see Table~\ref{table_properties_DRC_components}) we use their observed flux densities at $2 \, {\rm mm}$, corresponding to $\sim 400 \, {\rm \mu m}$ in the rest frame, which is the only wavelength where the continuum emission from all the DRC components has been detected.  The flux density at $2 \, {\rm mm}$ of each DRC component is converted to an SFR (see values for individual components in Table~\ref{table_properties_DRC_components}) by re-scaling the ALESS template and integrating it between rest-frame 8--$1000 \, {\rm \mu m}$.  This method assumes that the SEDs (from $8$ to $1000 \, {\rm \mu m}$ in the rest frame) of all DRC components are well represented by the ALESS template, a fact that should be considered when comparing DRC to other proto-clusters.  Under the same assumption, the continuum sensitivity of our ALMA $2 \, {\rm mm}$ observations indicates that we are sensitive to sources with ${\rm SFR} \gtrsim 40 \, M_\odot \, {\rm yr}^{-1} (3 \sigma)$. We include DRC-5 in Table~\ref{table_properties_DRC_components}, assuming that it belongs to the same proto-cluster structure, although no emission line has been detected from this component.  This is because our observations only covered $v > -1000 \, {\rm km \, s^{-1}}$ and, consequently, we cannot be sure that DRC-5 does not belong to the proto-cluster.

Among the SED templates used in \cite{Ivison2016ApJ...832...78I}, which include a representative range, only the one reported by \cite{Pearson2013MNRAS.435.2753P} yields lower SFRs than the ALESS SED template, by a factor $0.66\times$.  The Cosmic Eyelash SED \citep{Swinbank2010Natur.464..733S, Ivison2010A&A...518L..35I} and the SED template reported by \cite{Pope2008ApJ...675.1171P} yield similar SFRs to the ALESS template.  The Arp\,220 SED gives higher SFRs than the ALESS template by a factor $1.36\times$. The same happens for the SED of the lensed source, G15.141 at $z = 4.24$ \citep{Cox2011ApJ...740...63C, Lapi2011ApJ...742...24L}, which yields SFRs higher than the ALESS template by a factor $2.21\times$.  Therefore, although the uncertainties can be significant, the SED template used in this work to measure SFRs does not give the highest SFR -- we are being conservative (modulo the possibility of a profoundly different IMF in these objects -- \citealt{Romano2017MNRAS.470..401R}). The high SFRs measured for the DRC components are not artificially high because of the chosen template, but because they are truly luminous.

\begin{table}[!t]
\caption{\label{table_properties_DRC_components_line_fluxes}Line properties of DRC components}
\centering
\begin{tabular}{lccccccccc}
\hline
 Component & $I_{\rm [CI](1-0)}$ & $I_{\rm ^{12}CO(6-5)}$ & {\sc fwhm}$_{\rm ^{12}CO(6-5)}$\tablenotemark{(a)}	\\
  &  $[{\rm mJy \, km \, s^{-1}}]$ &$[{\rm mJy \, km \, s^{-1}}]$	& $[{\rm km \, s^{-1}}]$\\
\hline
DRC--1	&	$882 \pm 119$	&	$4192 \pm 331$	&	$1009 \pm 88$		\\	
DRC--2	&	$394 \pm 54$	&	$1748 \pm 337$	&	$2140 \pm 466$	\\	
DRC--3	&	$598 \pm 76$	&	$757 \pm 64$		&	$359 \pm 29$		\\	
DRC--4	&	$364 \pm 44$	&	$539 \pm 56$		&	$602 \pm 68$		\\	
DRC--5	&	--			&		--			&		--			\\	
DRC--6	&	--			&	$767 \pm 58$		&	$840 \pm 70$		\\	
DRC--7	&	--			&	$169 \pm 53$		&	$1296 \pm 472$	\\	
DRC--8	&	--			&	$130 \pm 24$		&	$515 \pm 104$		\\ 	
DRC--9	&	--			&	$109 \pm 18$		&	$380 \pm 70$		\\ 	
DRC--10	&	--			&	$52 \pm 14$		&	$288 \pm 85$		\\
DRC--11	&	--			&	$33 \pm 11$		&	$243 \pm 89$		\\
\hline
\tablenotetext{1}{Obtained from Gaussian fits to the line profiles. Note that the $^{12}$CO(6--5) emission in DRC-2 has a very high linewidth because of the boxy shape of the line profile.  In any case, the {\sc fwzi} of the $^{12}$CO(6--5) emission in DRC-2 is $\sim 1600 \, {\rm km \, s^{-1}}$.}
\end{tabular}
\end{table}

The total obscured SFR of DRC -- which, recall, is only the core of our larger over-density of DSFGs -- is as extreme as ${\rm SFR} \sim 6,500 \, {\rm {\rm M}_\odot \, yr^{-1}}$, and about 75\% of that is taking place in three DRC components: DRC-1, DRC-2 and DRC-3, those with $L_{\rm IR} \geq 5 \times 10^{12}\,L_\odot$.  The obscured SFR of the proto-cluster core derived from the ALMA continuum emission at $2 \, {\rm mm}$ is lower in a factor of $1.37 \times$ than that derived from the flux density associated with its extended LABOCA 870-$\mu$m emission (components A and B in the left panel of Fig.~\ref{figure_LABOCA_wide_map}), $S_{\rm 870 \mu m} = 64 \pm 11 \, {\rm mJy}$, which would imply ${\rm SFR} \sim 8,900 \, {\rm {\rm M}_\odot \, yr^{-1}}$.  It is also lower in a $1.34 \times$ factor than the obscured SFR associated to the total IR luminosity derived in \cite{Ivison2016ApJ...832...78I} from the {\it Herschel}+LABOCA/SCUBA-2 photometry.  One reason for this discrepancy could be that the ALESS SED template does not provide a good representation of the dust emission for all the DRC components, such that some of them would have higher SFRs than those reported in Table~\ref{table_properties_DRC_components}.

The total obscured SFR of our proto-cluster core is the highest reported so far for proto-clusters whose members have been spectroscopically confirmed at $z > 4$.  As a reference, the total obscured SFR of AzTEC-3, a massive proto-cluster at $z_{\rm spec} = 5.3$ \citep{Capak2011Natur.470..233C}, is ${\rm SFR} \sim 1,600 \, {\rm M}_\odot \, {\rm yr}^{-1}$ as estimated from the only DSFG in this system. For a comparison at slightly lower redshifts, the total obscured SFR in the core of the proto-cluster SSA22 at $z_{\rm spec} = 3.09$ is ${\rm SFR} \sim 3,820 \, M_\odot \, {\rm yr}^{-1}$, calculated from the observed flux density at $1.1 \, {\rm mm}$ of the components and adopting the ALESS SED template, as with DRC).  CL\,J001, a concentration of DSFGs at $z_{\rm spec} = 2.5$ \citep{Wang2016ApJ...828...56W} has ${\rm SFR} \sim 3,400 \, {\rm M}_\odot \, {\rm yr}^{-1}$ in the central $80 \, {\rm kpc}$ region, much lower than DRC.  One of the few comparable cases in terms of the high obscured SFR is the COSMOS proto-cluster at $z = 2.10$ reported by \citet{Yuan2014ApJ...795L..20Y,Hung2016ApJ...826..130H}, but its SFR was determined over much wider scales than the core of our proto-cluster and, additionally, the COSMOS structure is at much lower redshift than DRC.

As reported above, DRC is just the core of an over-density of DSFGs.  The total $870 \, {\rm \mu m}$ of all these DSFGs (the eight LABOCA sources, including DRC -- see left panel of Fig.~\ref{figure_LABOCA_wide_map}) is $\approx 106\,{\rm mJy}$.  If all those DSFGs were at the same redshift as DRC, the total SFR of the system would be around $14,400 \, {\rm M}_\odot \, {\rm yr}^{-1}$.  The fact that the DSFGs found around DRC are at its same redshift is supported by their photometric redshifts measured from the {\it Herschel} plus LABOCA/SCUBA-2 photometry (Lewis et al.\ in prep), although further observations are required to confirm this.

\begin{figure*}
\centering
\includegraphics[width=0.28\textwidth]{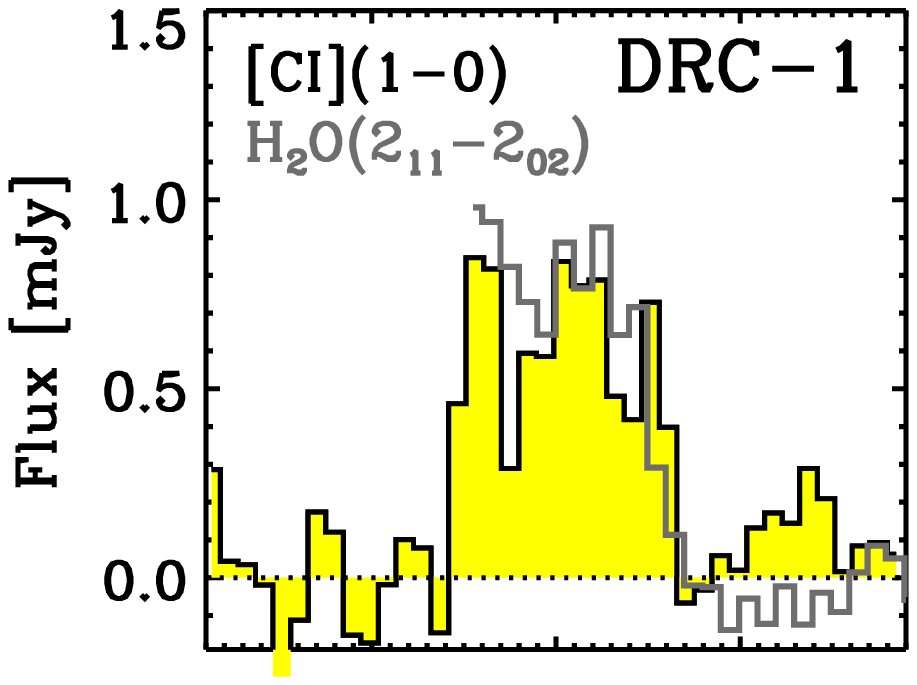}
\hspace{-16mm}
\includegraphics[width=0.28\textwidth]{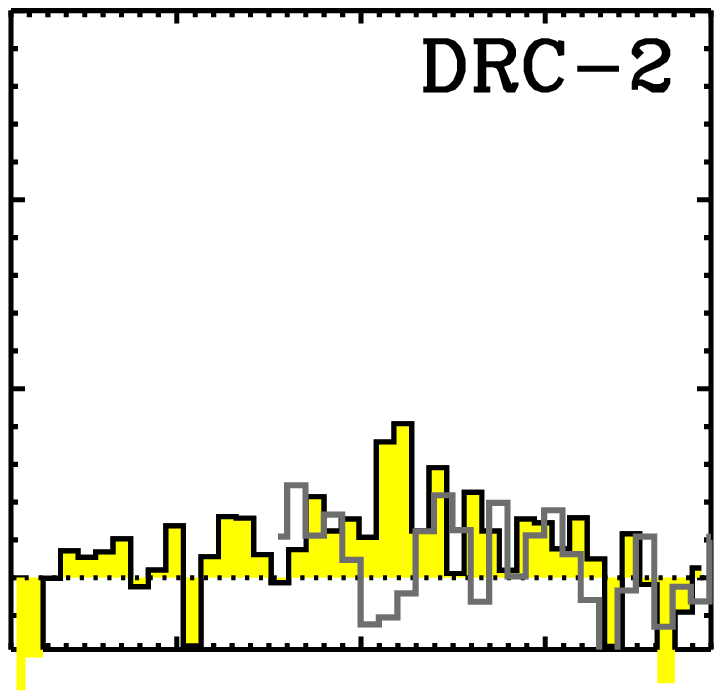}
\hspace{-16mm}
\includegraphics[width=0.28\textwidth]{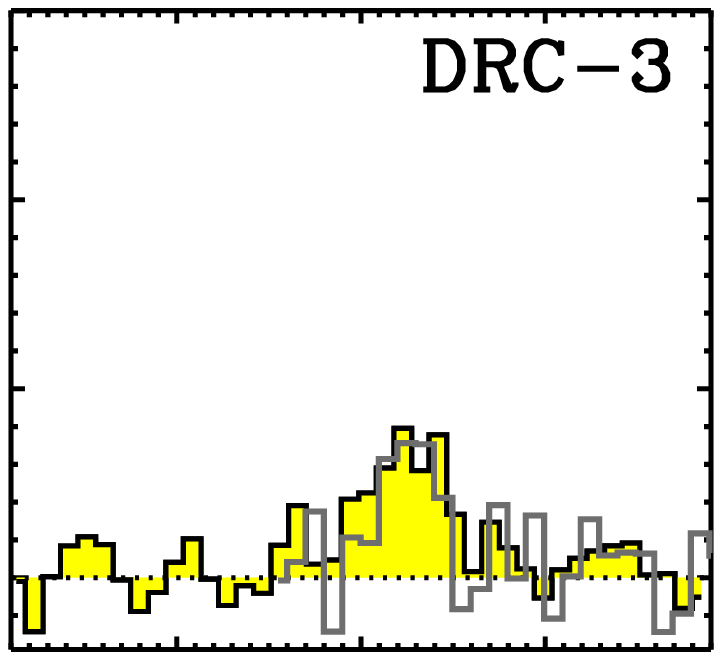}
\hspace{-16mm}
\includegraphics[width=0.28\textwidth]{empty.eps} \\

\vspace{-11mm}
\includegraphics[width=0.28\textwidth]{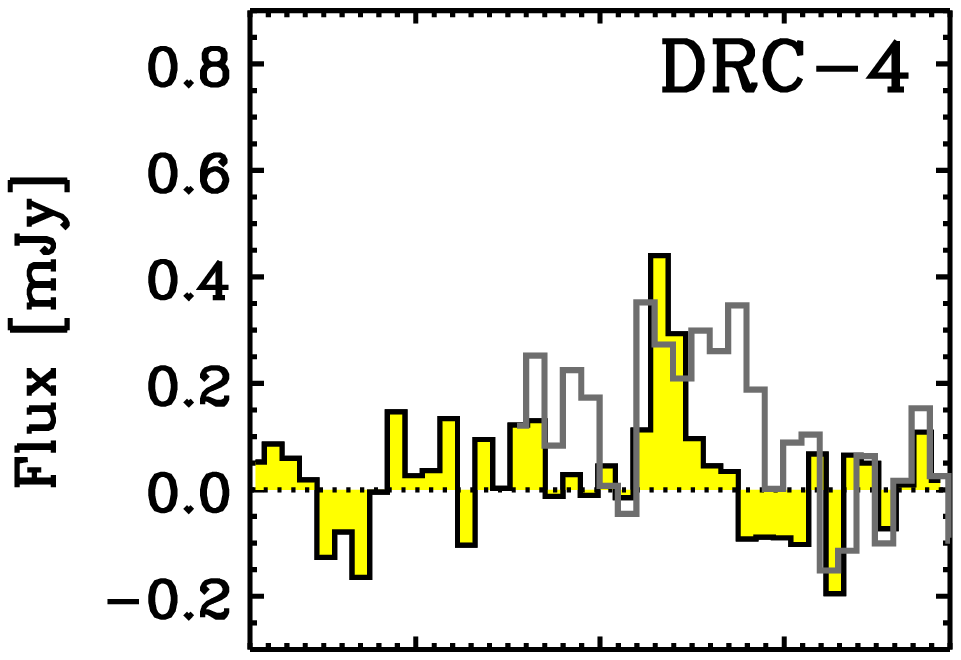}
\hspace{-16mm}
\includegraphics[width=0.28\textwidth]{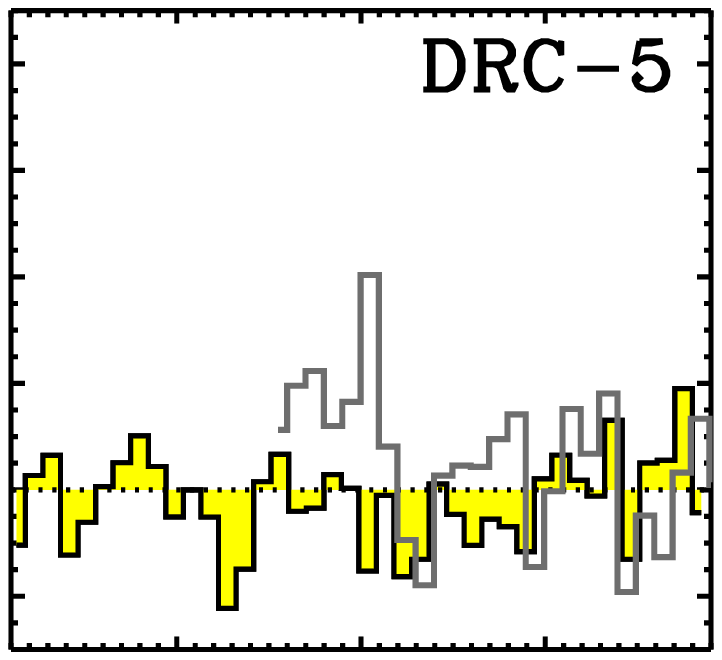}
\hspace{-16mm}
\includegraphics[width=0.28\textwidth]{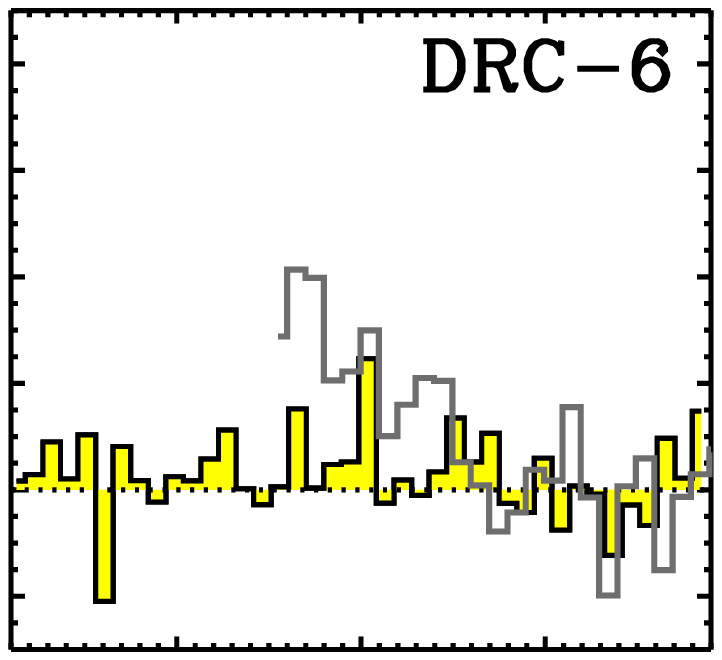}
\hspace{-16mm}
\includegraphics[width=0.28\textwidth]{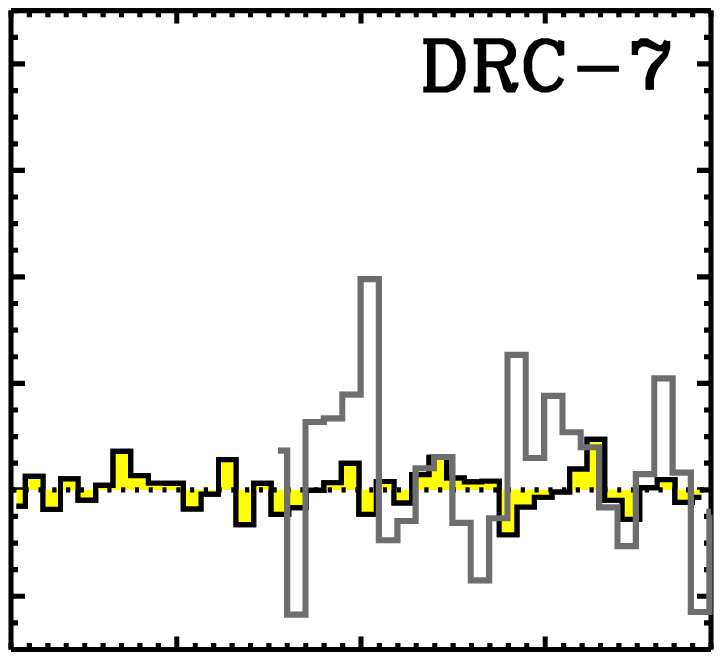} \\

\vspace{-11mm}
\includegraphics[width=0.28\textwidth]{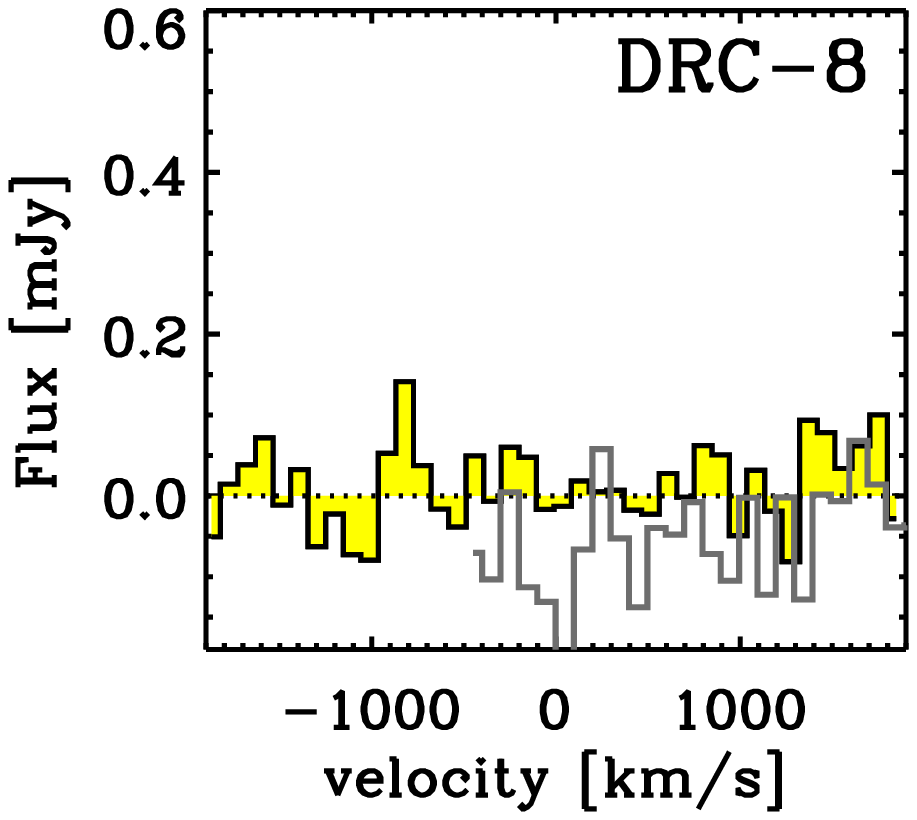}
\hspace{-16mm}
\includegraphics[width=0.28\textwidth]{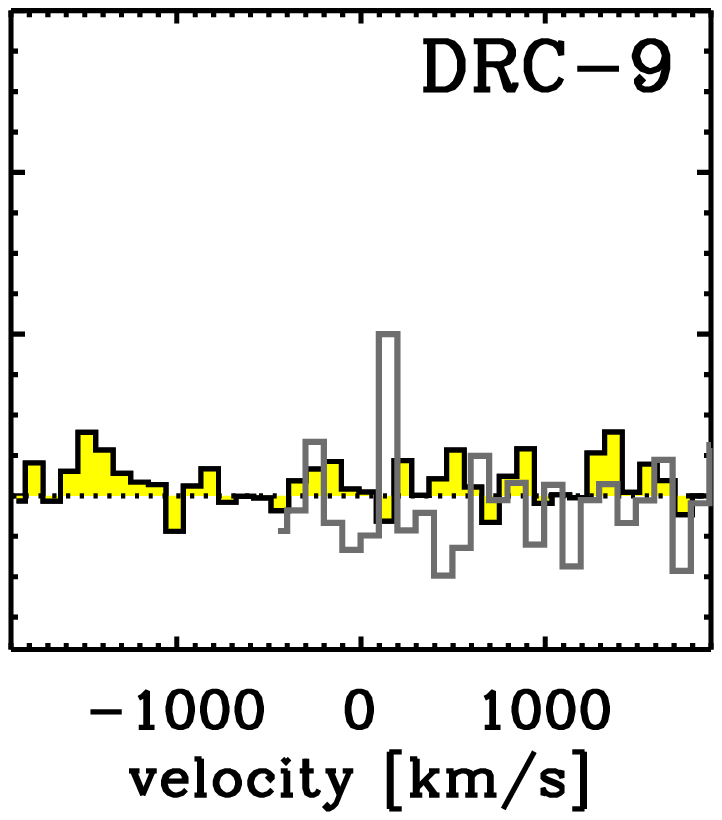}
\hspace{-16mm}
\includegraphics[width=0.28\textwidth]{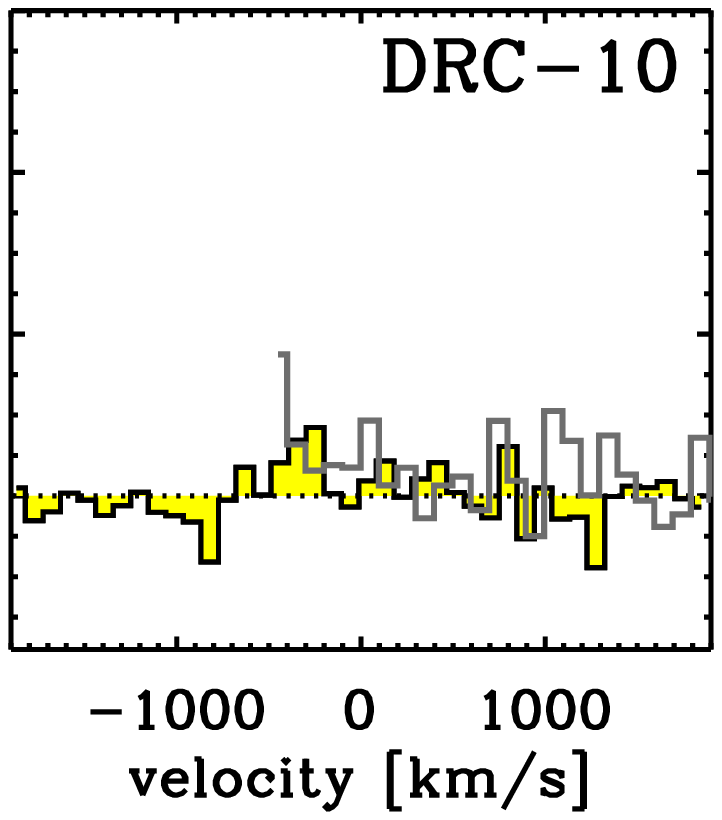}
\hspace{-16mm}
\includegraphics[width=0.28\textwidth]{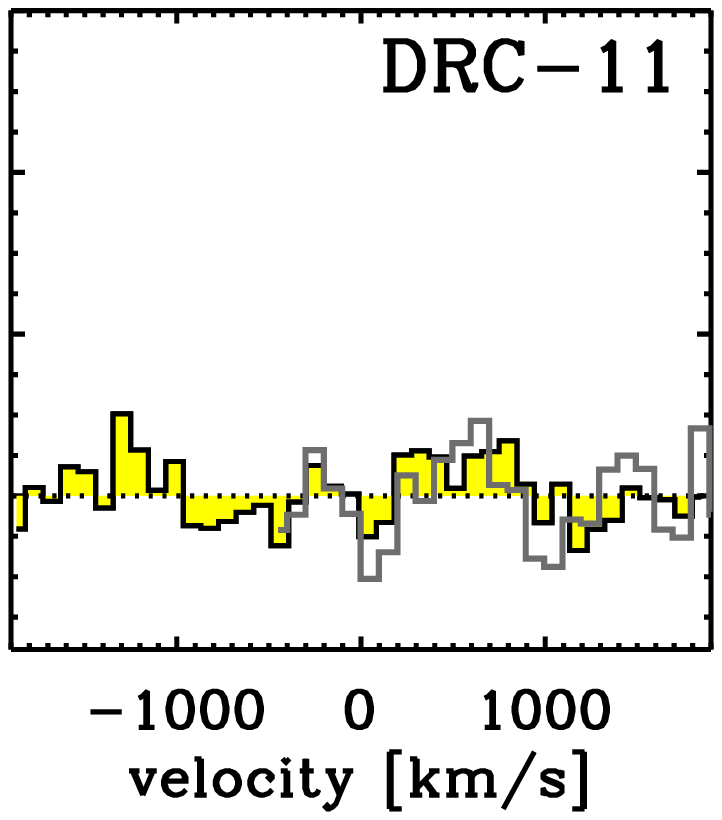} \\

\caption{ALMA continuum-subtracted [C\,{\sc i}](1--0) and ${\rm H_2O} (2_{11} - 2_{02})$ spectra of the 11 components in the core of our extreme proto-cluster, DRC, at $z_{\rm spec} = 4.002$.  The detection of these two additional emission lines in some DRC components provides unambiguous confirmation of the redshift.  Note that only about half of the line profile of the water emission has been detected in most of the DRC components.  This is because the spectral set-up of the $2 \, {\rm mm}$ observations was defined assuming that the emission line seen in our previous $3 \, {\rm mm}$ observations was CO; it was actually [C\,{\sc i}](1--0), so we did expect to cover the water transition.
\label{ALMA_spec_DRC_DDT}}
\end{figure*}

\subsection{Molecular gas mass} 

The [C\,{\sc i}](1--0) transition has been proposed to be a good tracer of the total molecular gas mass, even better than low-$J$ CO lines in high-redshift galaxies \citep{Papadopoulos2004MNRAS.351..147P} and it has been used to estimate the molecular gas of several populations of high-redshift DSFGs  \citep[e.g.][]{Walter2011ApJ...730...18W,Alaghband2013MNRAS.435.1493A,Bothwell2017MNRAS.466.2825B}. In this work we use [C\,{\sc i}](1--0) to estimate the molecular gas mass for the DRC components detected in that transition (see Fig.~\ref{ALMA_spec_DRC_DDT}).  The line fluxes have been derived from the moment-0 maps of [C\,{\sc i}](1--0) for each component. For the DRC components whose [C\,{\sc i}](1--0) line is not detected (see for example DRC-2, whose [C\,{\sc i}](1--0) is not clearly detected in the spectrum -- the line is expected to be extremely broad, judging by the $^{12}$CO(6--5) profile), we have assumed that the [C\,{\sc i}](1--0) and $^{12}$CO(6--5) transitions have the same width.  Then, we have calculated the ${\rm H_2}$ mass of each DRC component following \cite{Bothwell2017MNRAS.466.2825B} and \cite{Alaghband2013MNRAS.435.1493A}, but see also \cite{Papadopoulos2004ApJ...615L..29P,Papadopoulos2004MNRAS.351..147P,Weiss2003A&A...409L..41W,Weiss2005A&A...429L..25W}:

\begin{equation}
\label{equation_ci_mass}
\begin{split}
	M_{\rm H_2} = 1375.8 \, D_{\rm L}^2 \, (1+z)^{-1} \left( \frac{X_{\rm [C\,I]}}{10^{-5}} \right)^{-1} \\ 
	\left( \frac{A_{\rm 10}}{10^{-7} \, {\rm s}^{-1}} \right)^{-1} \times Q_{10}^{-1} \, S_{\rm [C\,I]} \, \Delta v
\end{split}
\end{equation}

\noindent
where $X_{\rm [C\,I]}$ is the [C\,{\sc i}] / ${\rm H_2}$ abundance ratio (assumed to be $X_{\rm [C\,I]} = 3 \times 10^{-5}$), $A_{10}$ is the Einstein coefficient ($A_{10} = 7.93 \times 10^{-8} \, {\rm s}^{-1}$) and $Q_{10}$ is the excitation factor, assumed to be $Q_{\rm 10} = 0.6$ \citep{Bothwell2017MNRAS.466.2825B}.  Equation \ref{equation_ci_mass} does not take into account the effect of the cosmic microwave background (CMB) on the [C\,{\sc i}](1--0) line strength \citep{daCunha2013ApJ...766...13D,Zhang2016RSOS....360025Z}. The CMB at $z = 4$ has a temperature of $T_{\rm CMB} \sim 13.7 \, {\rm K}$, which is not negligible compared to the upper level of the [C\,{\sc i}](1--0) line, $\sim 23.6 \, {\rm K}$. In order to estimate the magnitude of the effect of the CMB we take into account that the upper level of the [C\,{\sc i}] line lies between the upper level energies of $^{12}$CO(2--1) and $^{12}$CO(3--2).  Following \cite{daCunha2013ApJ...766...13D}, the effect of the CMB would then be a suppression of the velocity-integrated flux of the [C\,{\sc i}](1--0) line by a factor $\sim 2 \times$, assuming local thermodynamic equilibrium (the precise factor depends on the kinetic temperature). This means that the molecular gas masses of the DRC components would be a factor of $2 \times$ higher than those given by Equation~\ref{equation_ci_mass}.  This a very crude estimation, since the effect of the CMB on the observed line fluxes depends on a number of factors. The molecular gas masses of the DRC components, derived with the above assumptions, are quoted in Table~\ref{table_properties_DRC_components}. It can be seen that our proto-cluster is formed by galaxies with massive molecular gas reservoirs, whose gas masses range from $M_{\rm H_2} \sim 1.1 \times 10^{11} \, {\rm M}_\odot$ to $M_{\rm H_2} \sim 2.6 \times 10^{11} \, {\rm M}_\odot$. The total molecular gas is at least $M_{\rm H_2} \sim 6.6 \times 10^{11} \, {\rm M}_\odot$.

\begin{figure}[!t]
\centering
\includegraphics[width=0.45\textwidth]{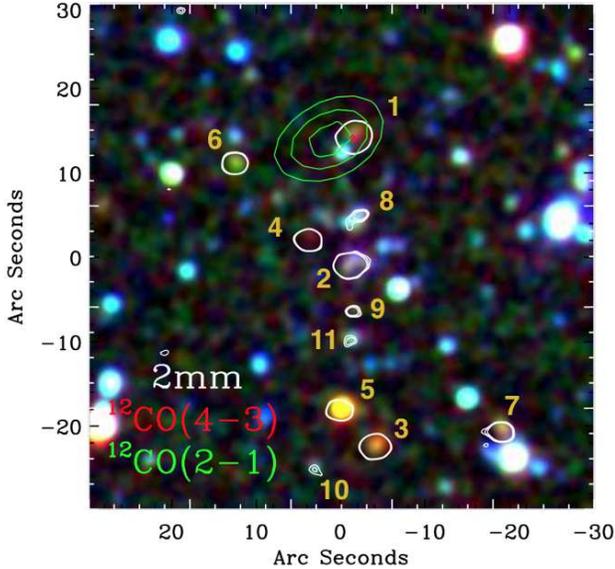} 
\caption{False-color image of the core of our proto-cluster at $z_{\rm spec} = 4.002$, obtained from our $I$-, $K_{\rm S}$-band and IRAC imaging.  The positions of the 11 DRC components are highlighted by white contours; $^{12}$CO(2--1) emission from ATCA is shown with green contours and the ALMA $^{12}$CO(4--3) emission is shown with red contours (note that this faint emission is on top of DRC-1).  All contours run from $4 \sigma$ to $6 \sigma$ in steps of $1 \sigma$. Most DRC components are detected in at least one optical/near-IR band and a variety of colors is seen, with the brightest sources at $2 \, {\rm mm}$ having the reddest optical/near-IR colors. }
\label{DRC_map_color_IRAC}
\end{figure}

\begin{figure}[!t]
\centering
\includegraphics[width=0.45\textwidth]{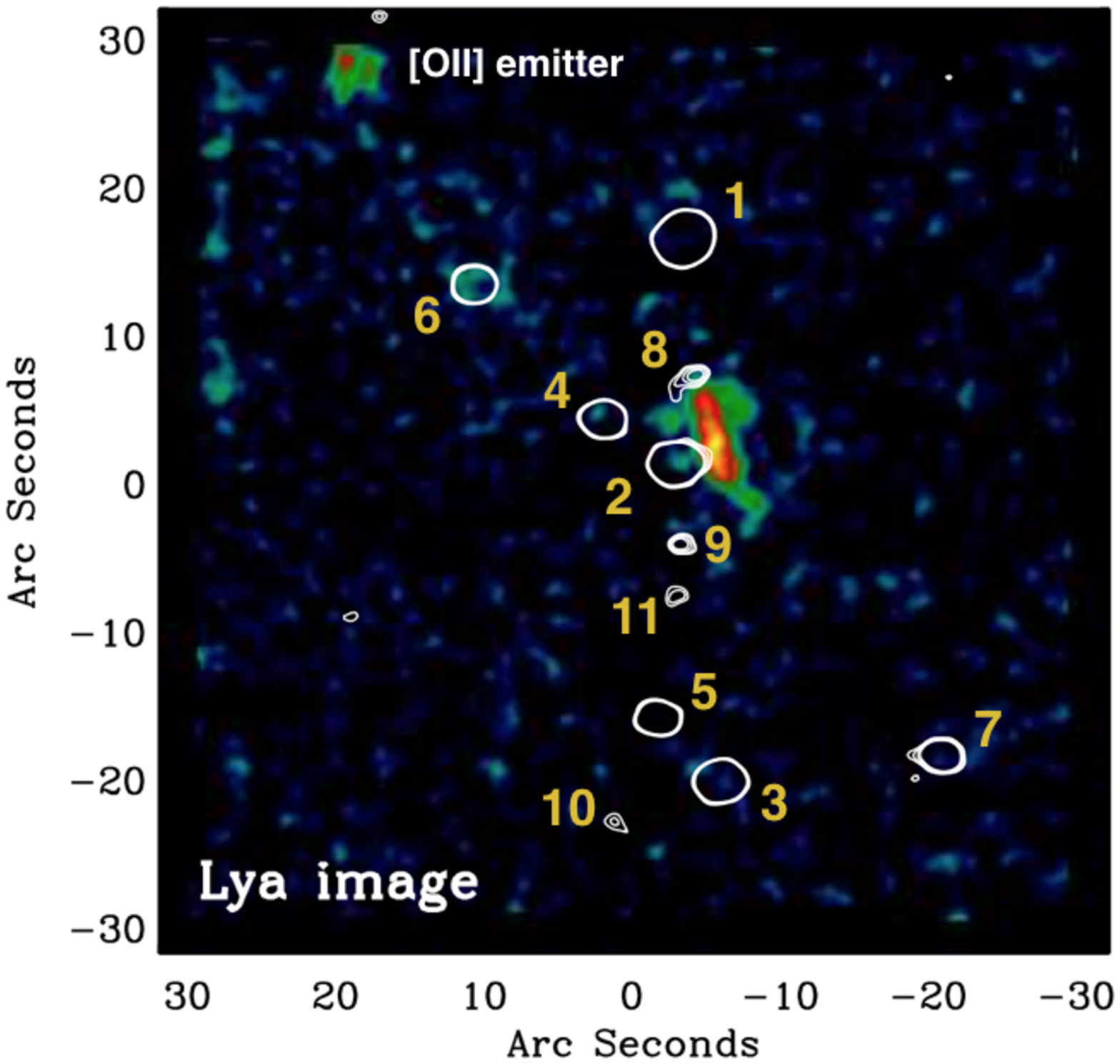} \\
\includegraphics[width=0.43\textwidth]{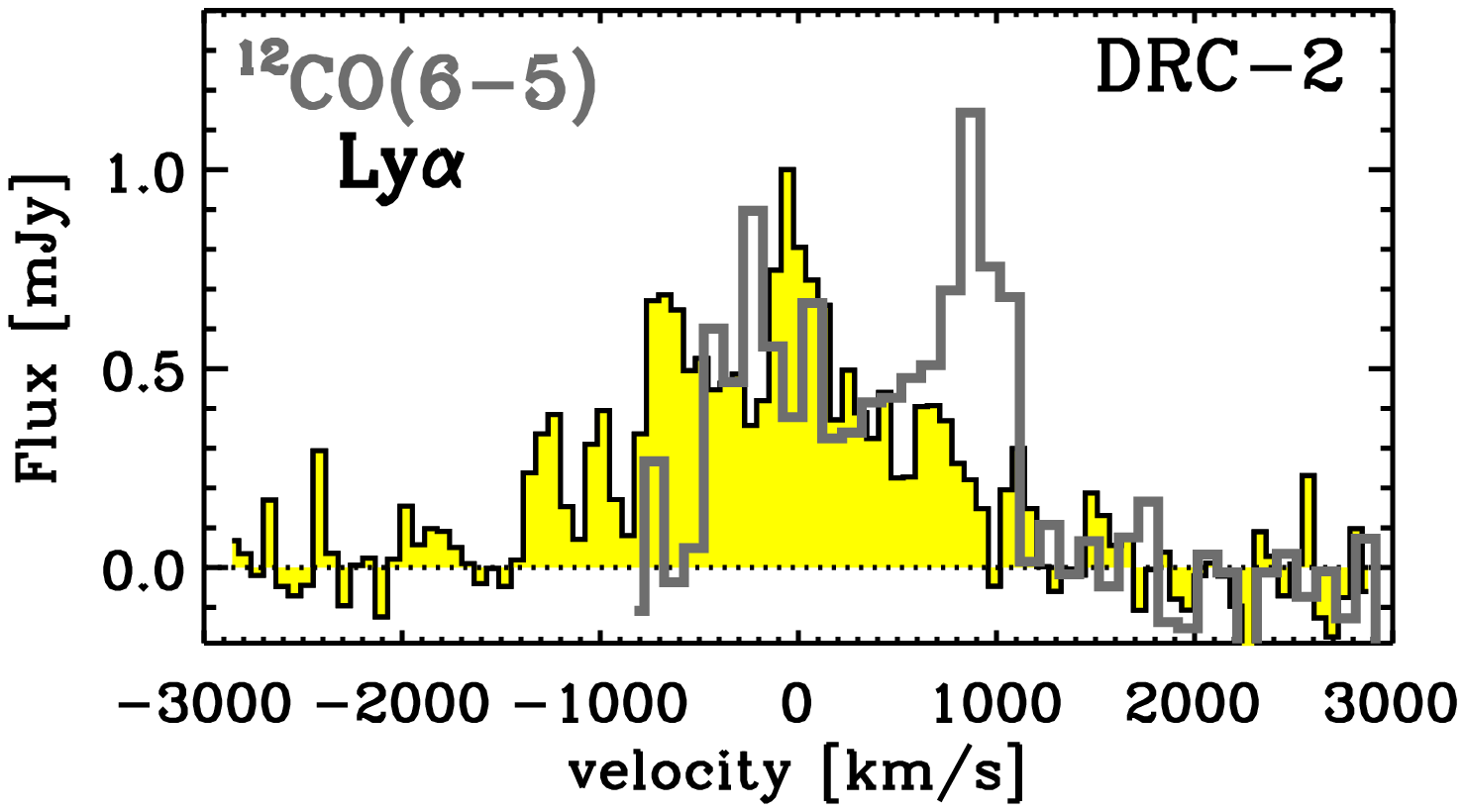} 
\caption{{\it Top}: Ly$\alpha$ image of the proto-cluster core, showing the Ly$\alpha$ blob next to DRC-2 (North is up; East is left) which extends over $60 \, {\rm kpc}$. Since the Ly$\alpha$ blob is faint, the image shown here has been smoothed using a $5 \, {\rm pixel} \times 5\,{\rm pixel}$ Gaussian kernel.  The emission seen to the north-east is [O\,{\sc ii}] from a star-forming galaxy at low redshift ($z_{\rm spec} = 0.637$).  The proto-cluster components are shown with white contours representing the $2 \, {\rm mm}$ continuum emission, which run from $4 \sigma$ to $6 \sigma$ in steps of $1 \sigma$.  Apart from the Ly$\alpha$ blob, no other Ly$\alpha$ emitter is detected in the field, indicating that the most extreme region of our proto-cluster is dominated by DSFGs and not by `normal' SFGs.  We note that some of the flux of the Ly$\alpha$ blob might be affected by the [O\,{\sc ii}] emission of a low-redshift SFG located at the southern end of the blob. {\it Bottom}: Spectrum of the Ly$\alpha$ blob, compared to the $^{12}$CO(6--5) emission in DRC-2.  The flux scale corresponds to the $^{12}$CO(6--5) transition, and the Ly$\alpha$ has been re-scaled to fit.  The Ly$\alpha$ blob profile is extremely broad (even broader than the $^{12}$CO(6--5) emission), with a FWZI greater than $2,000 \, {\rm km \, s^{-1}}$. This value is comparable to those seen in Ly$\alpha$ blobs associated with radio galaxies at $z \sim 4$ \citep{Swinbank2015MNRAS.449.1298S}.}
\label{DRC_Lya_image}
\end{figure}

Our ALMA $3 \, {\rm mm}$ scan also covered the $^{12}$CO(4--3) emission line. It was not detected because its faintness and because the noise was higher near its frequency. Once the redshift of DRC was confirmed, we calculated the moment-0 map of the $^{12}$CO(4--3) emission, assuming that it is as wide as the $^{12}$CO(6--5) transition detected in our ALMA band 4 observations. $^{12}$CO(4--3) emission is detected in the moment-0 map only from DRC-1.  Using this line flux to measure the molecular gas mass of this DRC component, assuming the average luminosity ratio of CO lines in DSFGs, $L'_{\rm CO(4-3)} / L'_{\rm CO(1-0)} = 0.46$ \citep{Ivison2011MNRAS.412.1913I,Bothwell2013MNRAS.429.3047B,Carilli2013ARA&A..51..105C} and the $\alpha_{\rm CO}$ for local ULIRGs, $\alpha_{\rm CO} = 0.8 {\rm M}_\odot \, ({\rm K \, km \, s^{-1} \, pc^2})^{-1}$ \citep{Downes1998ApJ...507..615D}, the derived molecular gas mass for DRC-1 from $^{12}$CO(4--3) is $M_{\rm H_2} \sim 1.1 \times 10^{11} \, {\rm M}_\odot$. This value is lower than the molecular gas mass of DRC-1 derived from [C\,{\sc i}](1--0), but is consistent given the large uncertainties involved in such determinations.

The molecular gas mass for the other DRC components could be determined from the detected $^{12}$CO(6--5) line transitions. However, this would lead to a very uncertain estimates because of the conversion from $^{12}$CO(6--5) to $^{12}$CO(1--0) luminosity and the conversion from $^{12}$CO(1--0) luminosity to molecular gas mass.  Alternatively, we could use the rest-frame $870 \, {\rm \mu m}$ luminosity as a proxy for the molecular gas mass, following \citet{Scoville2016ApJ...820...83S,Hughes2017arXiv170207350H,Oteo2017arXiv170705329O}.  For the DRC components with [C\,{\sc i}](1--0) detections, the molecular gas masses derived from [C\,{\sc i}](1--0) are $\sim 2-5 \times$ lower than those derived from the dust continuum luminosity, where the rest-frame luminosities at $870 \, {\rm \mu m}$ have been obtained by redshifting the ALESS template to $z = 4.002$ and re-scaling to the observed flux density of each DRC component at $2\, {\rm mm}$.  We report only the molecular gas mass of the sources detected in [C\,{\sc i}](1--0), noting that the total molecular gas mass of our proto-cluster might be much higher (also considering that some of the LABOCA-detected DSFGs around DRC might be at the same redshift).

\subsection{Gas-depletion time} 

We can estimate the gas-depletion time of the DRC components detected in [C\,{\sc i}](1--0) from the ratio between their molecular gas masses and their SFRs.  These are shown in Table~\ref{table_properties_DRC_components} and range between $\sim 90$--230$ \, {\rm Myr}$.  These gas-depletion times are comparable to those obtained for other luminous DSFGs at $z \gtrsim 4$ \citep{Riechers2013Natur.496..329R,Hodge2015ApJ...798L..18H,Oteo2016ApJ...827...34O,Fudamoto2017arXiv170708967F,Riechers2017arXiv170509660R}. 

The probability of detecting 10 short-lived, physically associated DSFGs -- we do not consider DRC-5 here due to the lack of spectroscopic confirmation -- is extremely low \citep[see for example][]{Casey2016ApJ...824...36C}, meaning that such structures are extraordinary systems.  The fact that some DRC components are detectable as luminous DSFGs at the same redshift and with relatively low gas-depletion times might suggest the existence of a mechanism able to trigger star formation simultaneously in different sources distributed across hundreds of ${\rm kpc}$.  Alternatively, given that the baryon to dark matter ratio in galaxies is much lower than the cosmic value, such that halos contain huge amounts of gas that can flow to the star-forming regions,  one can argue that star formation can be sustained over longer times until it is swept out by feedback effects, though the absence of `normal' SFGs is puzzling.  Relatively long starburst lifetimes have been reported in the past \citep{Granato2004ApJ...600..580G,Lapi2011ApJ...742...24L}.  

\subsection{Velocity dispersion and total mass}

Using the $^{12}$CO(6--5) line detections, which provide the highest signal to noise among all the detected lines, we have measured the central velocity of all the DRC components (see Table~\ref{table_properties_DRC_components}), from which we can derive a velocity dispersion, $\sigma_{\rm v} = 794 \, {\rm km \, s^{-1}}$. This relatively high velocity dispersion is mainly caused by the velocity of components DRC-7 and DRC-10 with respect to the others. The fact that these two components have velocity offsets as large as $\sim 2,000 \, {\rm km \, s^{-1}}$ with respect to the the average velocity of all the other components might mean that we are seeing two groups, as has been reported in proto-clusters at lower redshifts such as MRC\,0052-241 at $z = 2.86$ \citep{Venemans2007A&A...461..823V}.  We note that the velocity dispersion has been obtained from only 10 spectroscopically-confirmed sources in the core of the proto-cluster.

We can estimate the total mass of our proto-cluster using several different methods.  First, we can use the velocity dispersion ($\sigma_{\rm v}$) and the relation between the velocity dispersion and the total mass ($M_{\rm total}$) derived by \cite{Evrard2008ApJ...672..122E}, which has been also used to estimate the total mass of lower-redshift proto-clusters \citep{Wang2016ApJ...828...56W}:

\begin{equation}
	 \sigma_{\rm v} (M,z) = \sigma_{\rm DM,15} \left( \frac{h(z) M_{\rm total}}{10^{15} \, {\rm M}_\odot} \right)^\alpha
\end{equation}

\noindent
where $h(z)$ is the dimensionless Hubble parameter. Using $\sigma_{\rm DM, 15} \sim 1,083 \, {\rm km \, s^{-1}}$ and $\alpha \sim 0.336$ \citep[see also][]{Wang2016ApJ...828...56W} we derive a total mass of $M_{\rm total} \sim 9.3 \times 10^{13} \, {\rm M}_\odot$.  We note that this estimate is highly uncertain for two main reasons.  First, proto-clusters at $z \sim 4$ are not virialized.  Second, the velocity dispersion has been measured using only the spectroscopically-confirmed sources, which are all located in the core (the most violent region, which is forming stars at a rate of thousands of $M_\odot \, {\rm yr}^{-1}$). We can also estimate the total mass of the cluster by assuming a uniform spherical distribution with line-of-sight velocity dispersion and radius corresponding to those for our proto-cluster core.  In this case, the virial mass would be $M_{\rm total} \sim 3.2 \times 10^{13}\,M_\odot$, slightly lower than the previous estimate. Next, we estimate the total mass using the relation between the total IR luminosity and halo mass at $z = 4$ by \cite{Aversa2015ApJ...810...74A}, obtaining $M_{\rm total} \sim 4.4 \times 10^{13}\,M_\odot$.


The different methods used above to estimate the total mass of the proto-cluster core give a wide range of values, indicative of the uncertain nature of this task.  Despite the uncertainties, we have determined that our proto-cluster core is very massive. The total mass of the full proto-cluster would be considerably higher still, considering that DRC is only the core.  It could be even more massive than the most massive progenitor halos predicted by simulations.  As a reference, \cite{Chiang2013ApJ...779..127C} predicted $< 2 \times 10^{13} \, M_\odot$ at $z = 4$.  Using the evolutionary tracks of \cite{Chiang2013ApJ...779..127C}, we conclude that DRC might become an ultra-massive cluster at $z = 0$, with a total mass in the region of $2 \times 10^{15}\,M_\odot$; thus, DRC could be the progenitor of a Coma-like \citep{The1986AJ.....92.1248T,Kubo2007ApJ...671.1466K} or even more massive cluster.  However, we note that exploring the evolution of DRC is highly uncertain because we are only probing the core of a likely larger over--density of DSFGs and halos with similar masses can evolve very differently.  Further observations of the surroundings of DRC would be needed to have a better insight of the evolution (i.e. if DRC is truly the progenitor of a massive galaxy cluster or it is just a group of sources caught in a active stage during their evolution) of this extreme structure we have discovered.

\subsection{Stellar and Ly$\alpha$ emission}

Fig.~\ref{DRC_map_color_IRAC} shows a false-color image of the stellar emission in the components of our proto-cluster core, which has been created using $R$, $K_s$ and IRAC $4.5 \, {\rm \mu m}$ imaging.  A variety of colors can be seen, with the brightest components at $2 \, {\rm mm}$ being associated to reddest sources, as expected.  We note that our optical and near-IR imaging does not reveal any sign (like arcs) which might indicate that our proto-cluster is being gravitationally amplified by a foreground cluster.

Using our MUSE observations of the proto-cluster core we have created a continuum-subtracted Ly$\alpha$ image, which is shown in the top panel of Fig.~\ref{DRC_Lya_image}.  The Ly$\alpha$ image reveals the presence of Ly$\alpha$ blob next to DRC-2 and extended over $60 \, {\rm kpc}$, similar to what is found around others high-redshift DSFGs \citep{Ivison1998MNRAS.298..583I,Chapman2001ApJ...548L..17C,Umehata2015ApJ...815L...8U,Geach2016ApJ...832...37G} and radio galaxies \citep{Swinbank2015MNRAS.449.1298S}.  Actually, as discussed in \cite{Chiang2015ApJ...808...37C}, there is a tendency for extended Ly$\alpha$ emission in over-dense environments \citep{Matsuda2009MNRAS.400L..66M,Yang2009ApJ...693.1579Y,Erb2011ApJ...740L..31E,Matsuda2012MNRAS.425..878M}.  We note that part of the emission in this Ly$\alpha$ blob might be due to two low-redshift [O\,{\sc ii}] emitters located north and south the Ly$\alpha$ blob.  The  lower panel of Fig.~\ref{DRC_Lya_image} shows the spectrum of the Ly$\alpha$ blob, integrated over its full extent, in comparison to the $^{12}$CO(6--5) line profile of the DRC component right next to it.  The Ly$\alpha$ emission is even wider than the $^{12}$CO(6--5), but is still comparable to other Ly$\alpha$ blobs at these redshifts. 

It can be seen in Fig.~\ref{DRC_map_color_IRAC} that most of the DRC components detected at CO and $2 \, {\rm mm}$ continuum do not show significant Ly$\alpha$ emission, as expected for DSFGs (although there are exceptions to this -- \citealt{Chapman2003Natur.422..695C,Chapman2005ApJ...622..772C,Oteo2012ApJ...751..139O,Casey2012ApJ...761..139C,Oteo2012A&A...541A..65O,Sandberg2015A&A...580A..91S}). Furthermore, apart from the Ly$\alpha$ blob, no Ly$\alpha$ emitters are found in the $1' \times 1'$ central region of the proto-cluster (the emission-line galaxy detected in the north-east is a low-redshift [O\,{\sc ii}] emitter, with strong continuum and Balmer absorption lines), indicating that the star formation in this most extreme region is dominated by dust-obscured star formation. 

\section{Conclusions}
\label{conclusions_of_the_paper}

In this paper we have reported the identification of an extreme proto-cluster of DSFGs at $z_{\rm spec} = 4.002$ whose core (defined as the region where the SFR is maximal) is associated with one of the reddest sources in {\it H}-ATLAS (nicknamed DRC).  It comprises at least 10 DSFGs, distributed over an area of ${\rm 260 \, kpc \times 310 \, kpc}$, with a total SFR of at least $6,500 \, {\rm M}_\odot \, {\rm yr}^{-1}$. Most of this SFR is taking place in the dusty DSFGs -- our MUSE Ly$\alpha$ imaging reveals no {\it normal} star-forming galaxies in the proto-cluster core, just a $60 \, {\rm kpc}$-wide Ly$\alpha$ blob next to one of the DRC components, and at the same redshift.  The mass of the core of the proto-cluster is also extreme -- at least $\sim 6.6 \times 10^{11} \, M_\odot$, just in molecular gas.

DRC is the brightest component of an over-density of DSFGs discovered in a LABOCA wide-field observation around the proto-cluster core.  If all these additional DSFGs were at the same redshift as DRC -- meaning the total SFR would be $\sim 14,400 \, {\rm M}_\odot \, {\rm yr}^{-1}$ --  the structure would have an extent of at least ${\rm 2.3 \, Mpc \times 2.3 \, Mpc}$, close to the expected size of proto-clusters at $z \sim 4$ according to simulations, although the progenitors of the most massive clusters in the local Universe can extend across $\sim 15 \, {\rm Mpc}$ in diameter at the redshift of DRC \citep{Chiang2013ApJ...779..127C}.  

The core of the proto-cluster has a molecular gas mass of at least $6.6 \times 10^{11}\,M_\odot$ and its total mass could be as high as $\sim 4.4 \times 10^{13} \, M_\odot$. This is slightly more massive than models predict for the most massive progenitor halos and could suggest that DRC may evolve into a cluster at $z = 0$ with a total mass $> 10^{15} \, M_\odot$ and, therefore, could be the early progenitor of a cluster at least as massive as Coma.

The gas-depletion times of those DRC components with reliable molecular gas mass determinations are relatively low, ranging between 90 and $230\,{\rm Myr}$. This suggests either the presence of an unknown mechanism, able to trigger extreme star formation almost simultaneously in sources distributed over a few hundred ${\rm kpc}$ scales in the early Universe, or alternatively the presence of gas flows from the cosmic web, able to sustain star formation over much longer times than the estimated gas depletion times.

\bibliographystyle{mn2e}
\bibliography{cii_references}

\acknowledgments

IO, RJI, LD, SM, Z-YZ and AJRL acknowledge support from the European Research Council (ERC) in the form of Advanced Grant, {\sc cosmicism}. IO acknowledges G.~Bendo and M.~Zwaan for their help with the ALMA data calibration and analysis. LD also acknowledges support from ERC Consolidator Grant, CosmicDust. DR acknowledges support from the National Science Foundation under grant number AST-1614213. We would like to thank M.~J.~Micha{\l}owski for comments on the paper. This paper makes use of the following ALMA data: ADS/JAO.ALMA\#2013.1.00449.S, ADS/JAO.ALMA\#2013.A.00014.S, ADS/JAO.ALMA\#2013.A.00014.S, 2013.1.00001.S and 2016.1.01287.S. ALMA is a partnership of ESO (representing its member states), NSF (USA) and NINS (Japan), together with NRC (Canada) and NSC and ASIAA (Taiwan) and KASI (Republic of Korea), in cooperation with the Republic of Chile. The Joint ALMA Observatory is operated by ESO, AUI/NRAO and NAOJ. The National Radio Astronomy Observatory is a facility of the National Science Foundation operated under cooperative agreement by Associated Universities, Inc. Based on observations made with ESO Telescopes at the La Silla Paranal Observatory under programmes ID 295.A-5029 and ID 093.A-0705.  The Australia Telescope Compact Array is part of the Australia Telescope National Facility which is funded by the Australian Government for operation as a National Facility managed by CSIRO.  Based on observations obtained at the Gemini Observatory, which is operated by the Association of Universities for Research in Astronomy, Inc., under a cooperative agreement with the NSF on behalf of the Gemini partnership: the National Science Foundation (United States), the National Research Council (Canada), CONICYT (Chile), Ministerio de Ciencia, Tecnolog\'{i}a e Innovaci\'{o}n Productiva (Argentina), and Minist\'{e}rio da Ci\^{e}ncia, Tecnologia e Inova\c{c}\~{a}o (Brazil). This work is based [in part] on observations made with the Spitzer Space Telescope, which is operated by the Jet Propulsion Laboratory, California Institute of Technology under a contract with NASA.

\end{document}